\documentclass{article}
\usepackage{amssymb,multirow} 
\usepackage{amsmath}
\usepackage{cite} 
\usepackage[dvips]{graphicx,epsfig}
\usepackage{graphics}
\usepackage{setspace}
\def\1ad{\mbox{\normalsize $^1$}}
\def\2ad{\mbox{\normalsize $^2$}}
\def\3ad{\mbox{\normalsize $^3$}}
\def\4ad{\mbox{\normalsize $^4$}}
\def\5ad{\mbox{\normalsize $^5$}}
\def\6ad{\mbox{\normalsize $^6$}}
\def\7ad{\mbox{\normalsize $^7$}}
\def\8ad{\mbox{\normalsize $^8$}}

\makeatletter
\@addtoreset{equation}{section}
\makeatother

\def\beq{\begin{equation}}                     %
\def\eeq{\end{equation}}                       %
\def\bea{\begin{eqnarray}}                     
\def\eea{\end{eqnarray}}                       
\def\0 {\nonumber}

\def\a{\alpha}

\begin{document}

\setcounter{page}{0}
\begin{titlepage}
\titlepage
\rightline{SISSA 22/2008/FM}
\vskip 3cm
\centerline{{ \bf \Large Exact results for topological strings on resolved $Y^{p,q}$ singularities}}
\vskip 1.5truecm
\centering
{\bf Andrea Brini and Alessandro Tanzini}
\vskip 1.5cm
\begin{center}
\em 
 Mathematical Physics sector \\
International School for Advanced Studies (SISSA/ISAS) \\ 
Via Beirut 2, I-34014, Trieste, Italy\\
and Istituto Nazionale di Fisica Nucleare (INFN), sezione di Trieste

\vskip 3cm
\end{center}
\begin{abstract}
We obtain exact results in $\alpha'$ for open and closed $A-$model topological string amplitudes 
on a large class of toric Calabi-Yau threefolds by using their correspondence
with five dimensional gauge theories. The toric Calabi-Yaus that we analyze
are obtained as minimal resolution of cones over $Y^{p,q}$ manifolds
and give rise via M-theory compactification to $SU(p)$ gauge theories
on $\mathbb{R}^4\times S^1$. 
 As an application we present a detailed study of the local $\mathbb{F}_2$ case and compute open and closed genus zero Gromov-Witten invariants of the $\mathbb{C}^3/\mathbb{Z}_4$ orbifold. We also display the modular structure  of the topological wave function and give predictions for higher genus amplitudes.  The mirror curve in this case is the spectral curve of the relativistic $A_1$ Toda chain.
Our results also indicate the existence of a wider class of relativistic integrable systems associated to generic $Y^{p,q}$ geometries. \\

\vskip2cm

\end{abstract}

\vskip1.5\baselineskip

\vfill
 \hrule width 5.cm
\vskip 2.mm
{\small 
\noindent }
\begin{flushleft}
brini@sissa.it, tanzini@sissa.it
\end{flushleft}
\end{titlepage}
\large
\tableofcontents
\section{Introduction}

Since their formulation, 
topological theories have been a most fruitful 
source of results and ideas both in physics and mathematics. 
Topological amplitudes naturally arise in the BPS sector of superstrings \cite{agnt,bcov}
and supersymmetric gauge theories \cite{tym} and as such have a wide range of applications, from the evaluation of BPS protected terms in low-energy effective actions
to black hole microstates counting \cite{osv}.
Moreover, topological theories have provided new
and powerful tools for the computation of global properties of
manifolds,
{\it e.g.} Donaldson polynomials, Gromov-Witten invariants,
revealing at the same time surprising relationships
between
seemingly very different areas of mathematics. 

One of the most appealing features of topological strings is 
that the calculation of its amplitudes can be pushed to high
orders, sometimes to {\it all orders}, in perturbation theory.
To this end, one exploits symmetries and recursion relations
coming either from the underlying
${\cal N}=(2,2)$ supersymmetric sigma-models - as mirror symmetry
and holomorphic anomaly equations \cite{bcov} - or from the 
properties of some specific class of target manifolds - as
localization and geometric transitions for A-model on toric Calabi-Yaus  
\cite{topv}, or $\mathcal{W}$-algebras and integrable hierarchies on
the corresponding $B$-model side \cite{Aganagic:2003qj}.
 
These methods have been mostly applied in
the large volume region of the Calabi-Yau, where the
perturbative expansion in $\alpha'$ is well-behaved
and the topological string partition function has a clear
geometric interpretation as generating functional of
Gromov-Witten invariants. 
However, away from the large volume region, the perturbative
series diverges and the corresponding geometrical interpretation
breaks down. 
Very few exact results are known outside this perturbative regime,
although
significant progress has been recently obtained by using modular invariance
\cite{Aganagic:2006wq} and new matrix-model inspired techniques \cite{Bouchard:2007ys}.

In this paper, we obtain exact results in $\alpha'$ for a large class of 
toric Calabi-Yau threefolds, and calculate the corresponding topological
string amplitudes in the full moduli space of closed and open strings.
The basic idea is to resort to the correspondence with five-dimensional
gauge theories via M-theory compactification on the Calabi-Yau times
a circle \cite{Lawrence:1997jr}. More precisely, the geometries that we consider are obtained
from minimal resolution of $Y^{p,q}$ singularities, and M-theory compactification
over them give rise to $SU(p)$ gauge theories on $\mathbb{R}^4\times S^1$ with a $q$-dependent five dimensional Chern-Simons term \cite{Intriligator:1997pq,Hollowood:2003cv}. 
The mirror geometry can be written as a fibration
over an hyperelliptic curve, whose periods provide a
basis for the solutions of the $B$-model Picard-Fuchs equations.
Our main result is that we get
a closed form for the (derivatives of the)
periods on the whole $B$-model moduli space.
We then expand them in different patches
and calculate in this way topological amplitudes not only in the large volume region,
but in all phases, including orbifold and conifold points. 
The analytic continuation properties and modular structure underlying higher genus
amplitudes can be easily worked out in our approach.
As an application, we give predictions for 
Gromov-Witten invariants for the orbifold $\mathbb{C}^3/\mathbb{Z}_4$,
which corresponds to the blow-down of local $\mathbb{F}_2$ geometry ($p=q=2$).

We observe that, for $p=q$, the hyperelliptic curve appearing in the mirror geometry 
can be identified with the spectral curve of an integrable system, given by the relativistic generalization
of the $A_{p-1}$ Toda chain \cite{ruijsenaars}.   
The fact that with our method we can find closed formulae  
for any value of the parameter $q$ 
suggests the existence of a wider class of relativistic integrable systems.

The structure of the paper is the following: in Section 2 we review the toric geometry
of $Y^{p,q}$ singularities and their minimal resolutions, in Section 3 we discuss
mirror symmetry and the relation with integrable systems,
in Section 4 we outline our procedure to find topological amplitudes in the whole 
B-model moduli space, in Section 5 we provide some 
preliminary checks of our formalism.
In Section 6 we present a detailed study of the local $\mathbb{F}_2$ case:
we first compute 
open and closed genus zero Gromov-Witten invariants 
of the $\mathbb{C}^3/\mathbb{Z}_4$ orbifold, then analyze the modular 
properties of the topological wave function and use them to predict higher genus invariants.
We conclude in Section 7 with some comments and future perspectives.
Some technical details on the analytic continuation
of topological amplitudes are collected in the Appendix.
   
\section{Cones over $Y^{p,q}$}
\label{sectypq}

The toric geometry of $Y^{p,q}$ singularities \cite{Gauntlett:2004yd} has been extensively studied in the context of AdS/CFT correspondence \cite{Martelli:2004wu},
with the aim to provide non-trivial checks\footnote{See also \cite{bertolini} for related work.} for superconformal theories with reduced amount of
supersymmetry. We observe here that minimal resolution of such singularities gives rise
precisely to the local Calabi-Yau geometries that one usually considers to ``geometrically engineer'' gauge theories
via M-theory compactifications \cite{Hollowood:2003cv}.
   
The manifolds $Y^{p,q}$, with $p$ and $q$ integers such that $1<q<p$, are an infinite class of five-dimensional manifolds on
which explicit Sasaki-Einstein metrics can be constructed \cite{Gauntlett:2004yd}; 
the two extremal cases $q=0$ and $q=p$ may be formally added to the family, corresponding  
to $\mathbb{Z}_p$ quotients respectively of $T^{1,1}$ (the base of the conifold) and of $S^5/\mathbb{Z}_2$. 
Since $Y^{p,q}$ are Sasaki-Einstein, the metric cone $C(Y^{p,q})$ constructed over them is K\"ahler Ricci-flat; 
moreover, given that the base has a $\mathbb{T}^3$ of isometries effectively acting by (Hamiltonian) symplectomorphisms, 
the cone is a \textit{toric} threefold \cite{Martelli:2004wu}, that is, it contains an algebraic three-torus 
$(\mathbb{C}^*)^3$ as a dense open subset acting on the full variety through an extension of the natural action on 
itself (for an introduction to toric geometry see for example \cite{fulton, Bouchard:2007ik}).
As any toric $CY$ threefold, its geometry is fully codified by a three dimensional 
fan $\Sigma$ whose rays end on an affine hyperplane, say $r_3=1$, in the three dimensional space $\mathbb{R}^3$
with coordinates $(r_1,r_2,r_3)$. For $C(Y^{p,q})$, this is 
given by the following four lattice vectors in $\mathbb{Z}^3$:

\beq
v_1=\left(
\begin{array}{c}
1\\0\\1
\end{array}
\right),
\quad
v_2=\left(
\begin{array}{c}
0\\0\\1
\end{array}
\right),
\quad
v_3=\left(
\begin{array}{c}
0\\p\\1
\end{array}
\right),
\quad
v_4=\left(
\begin{array}{c}
-1\\p-q\\1
\end{array}
\right)
\eeq
In the following we will be interested in investigating the ($GKZ$ extended) K\"ahler moduli space of toric and 
canonical class preserving complete resolutions of $C(Y^{p,q})$. At the level of the toric diagram, 
this amounts \cite{Bouchard:2007ik} to add the $p-1$ internal points $v_{4+j} = (0,j,1)$ for $j=1,\dots,p-1$ and 
declare that the set of three dimensional cones in the fan $\Sigma$ is given by the simplicial cones whose 
projection on the $r_3=1$ hyperplane yields a triangulation of the polyhedron $\{ v_1,v_2,v_3,v_4 \}$. 

\begin{figure}[!h]
\vspace{1cm}
\begin{minipage}[t]{0.49\linewidth}
\centering
\vspace{0pt} 
\includegraphics[scale=0.5]{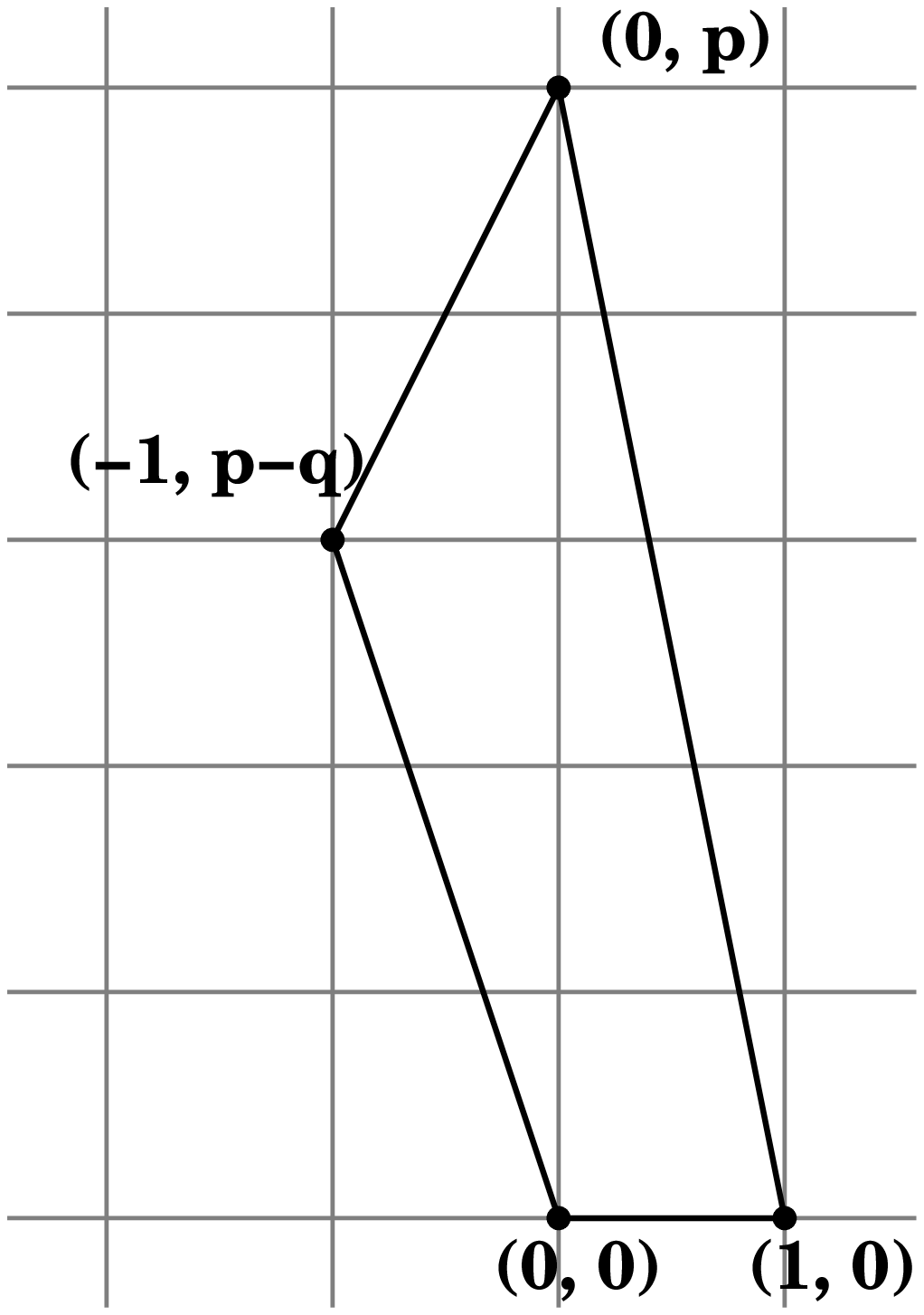}
\vspace{0pt} 
\caption{The fan of $C(Y^{p,q})$ for $p=5$, $q=2$.}
\label{fanypqsing}
\end{minipage} 
\begin{minipage}[t]{0.49\linewidth}
\centering
\vspace{0pt} 
\includegraphics[scale=0.5]{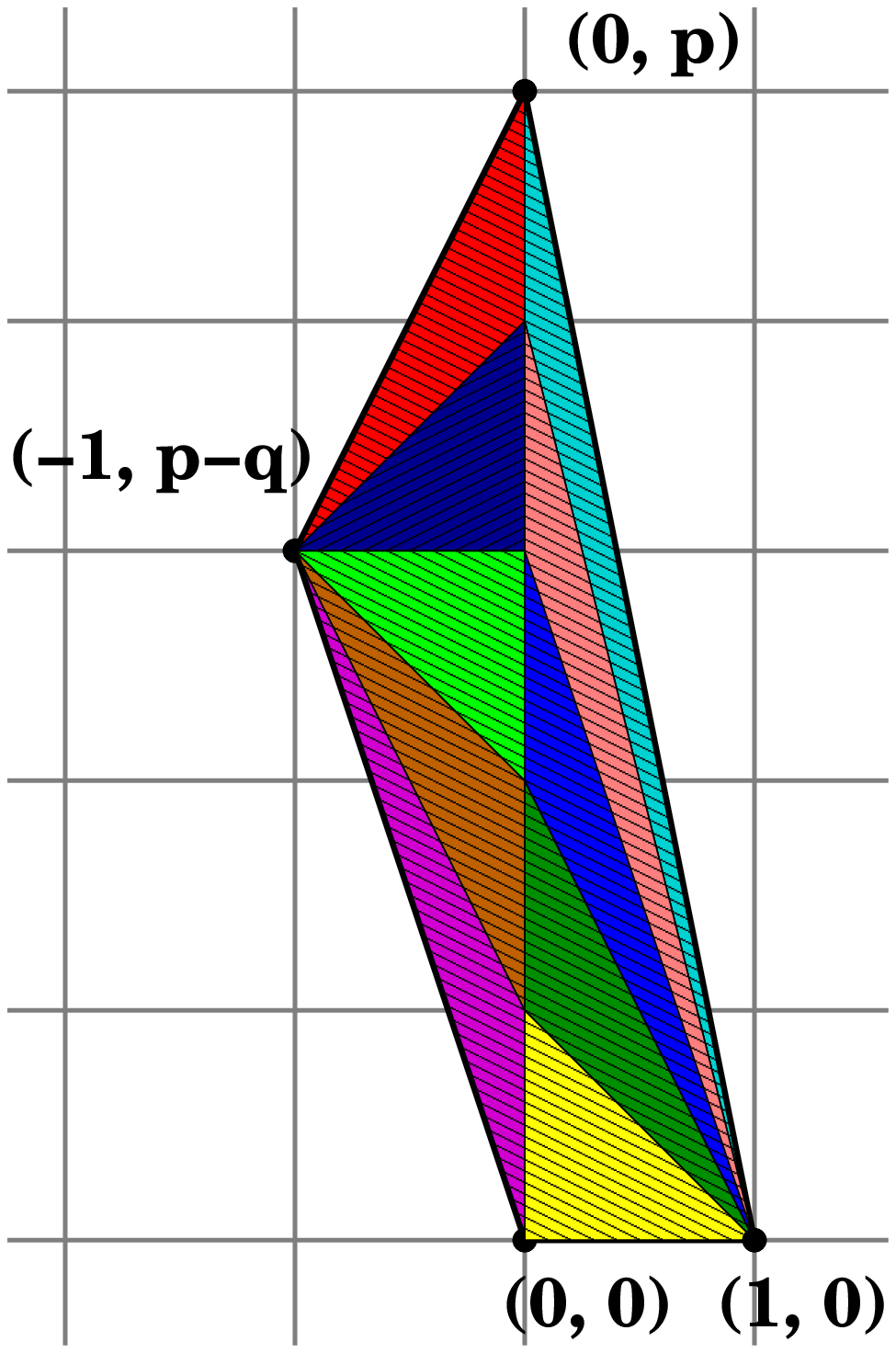}
\vspace{0pt} 
\caption{The fan of $\widetilde{C(Y^{p,q})}$ for $p=5$, $q=2$.}
\label{fanypqres}
\vspace{1.5cm}
\end{minipage} 
\end{figure}
\begin{figure}[b]
\centering
\includegraphics[scale=1]{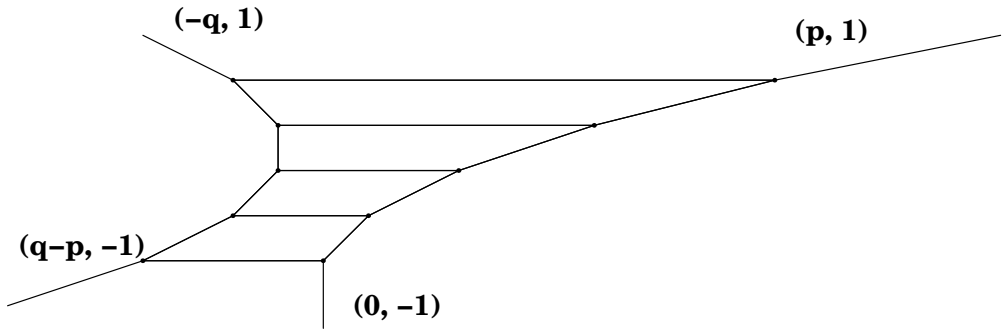}
\caption{The $pq$-web diagram of $\widetilde{C(Y^{p,q})}$ for $p=5$, $q=2$.}
\label{pqweb}
\end{figure}

The resolved geometry will be henceforth denoted as $\mathcal{X}_{p,q} \equiv \widetilde{C(Y^{p,q})}$
and the corresponding fan as $\Sigma_{p,q}$. It might also be described as a holomorphic quotient
$$(\mathbb{C}^{p+3} \setminus Z)/(\mathbb{C}^*)^p$$
with $Z$ a co-dimension $>0$ locus determined by the toric data \cite{fulton} and the $k^{th}$ $\mathbb{C}^*$ factor acting on the coordinates of $\mathbb{C}^{p+3}$ as $z_i \to \lambda^{Q^{(k)}_i} z_i$, where $Q_i^{(k)}\in \mathbb{Z}$ is a set of integers such that

\beq
\sum_{i=1}^{p+3}Q_i^{(k)} v_i =0 \qquad  k=1,\dots,p
\eeq
The set of charges for $\mathcal{X}_{p,q}$ is given by (see fig. \ref{fanypqres})
\beq
\label{glsmpq}
\begin{array}{ccccccccccccc}
Q_1 &=& (A, & - 2 A - B, & B, & A, &, 0, & 0, & 0, & 0, & 0, & 0, & 0) \\
Q_2 &=& (0, & 1, & 0, & 0, & -2, & 1, & 0, & 0, & \dots, & 0, & 0) \\ 
Q_3 &=& (0, & 0, & 0, & 0, & 1, & -2, & 1, & 0, & \dots, & 0, & 0) \\
Q_4 &=& (0, & 0, & 0, & 0, & 0, & 1, & -2, & 1, & \dots, & 0, & 0) \\
\vdots & & \vdots & \vdots &\vdots &\vdots &\vdots &\vdots &\vdots &\vdots &\vdots &\vdots & \vdots \\ 
Q_p &=& (0, & 0, & 1, & 0, & 0, & 0, & 0, & 0,  & \dots, & 1, & -2)
\end{array}
\eeq
with $A$ and $B$ coprime numbers solving the Diophantine equation $(p-q) A + p B = 0$ for $q<p$, while $A=1$ and $B=0$ for $q=p$. In real polar coordinates ($|z_i|, \theta_i$), this corresponds to the Higgs branch of a $\mathcal{N}=(2,2)$ $d=2$ gauged linear $\sigma$-model 
(GLSM) \cite{Witten:1993yc} with $p+3$ chiral fields $z_i$. The $D-$term equation of motion is
\beq
\label{dtermglsm}
\sum_{j=1}^{p+3} Q_j^{k} |z_i|^2 = t_k \qquad k=1, \dots, p
\eeq
and $U(1)^p$ acts as
\beq
\label{gfixglsm}
z_j \to e^{2 \pi i Q_j^{(k)} \theta_j} z_j  \qquad k=1, \dots, p
\eeq
where $\theta_j=\hbox{arg}(z_j)$. The Fayet-Iliopoulos parameters $t_k$ are complexified K\"ahler parameters of $\mathcal{X}_{p,q}$. Indeed, the full cohomology ring of the smooth $CY$ manifold thus obtained can be easily read off from the fan \cite{fulton,Benvenuti:2004dy}. For example, Betti numbers are
\beq
\label{coomologiaypq}
b_0=1 \qquad b_2=p \qquad b_4=p-1   \qquad b_6=0
\eeq
Various aspects of 
these geometries have been considered in the context of topological strings. First of all, notice from figure \ref{fanypqsing}-\ref{pqweb}   and formula (\ref{coomologiaypq}) that for $p=1$ we encounter the two most studied local curves: the conifold $(q=0)$ and $\mathbb{C} \times K_{\mathbb{P}^1}$ ($q=1$), whose enumerative geometry \cite{faberpanda}, phase structure \cite{Caporaso:2006gk} and local mirror symmetry properties \cite{Forbes:2005xt} have been extensively studied. For $p=2$, the local $CY$ in question is the total space of the canonical line bundle $K_{\mathbb{F}_q}$ over the $q^{th}$ Hirzebruch surface, $q=0,1,2$. For higher $p$ we have the ladder geometries considered in \cite{Katz:1996fh, Iqbal:2003zz, Eguchi:2000fv, Hollowood:2003cv} in the context of geometric engineering of pure $SYM$ theories with eight supercharges. In a suitable field theory limit described in \cite{Katz:1996fh}, the Gromov-Witten large radius expansion for these geometries was shown to
  reproduce for all $q$ the weak coupling instanton expansion of the prepotential for $\mathcal{N}=2$ $SU(p)$ pure Yang-Mills in $d=4$. Subsequently, they were shown \cite{Intriligator:1997pq, Lawrence:1997jr, Tachikawa:2004ur}, to geometrically engineer  $\mathcal{N}=1$ $SU(p)$ $SYM$ on $\mathbb{R}^4 \times S^1$ with Chern-Simons coupling $k=p-q$, with the field theory limit above interpreted now as the limit in which the fifth-dimensional circle shrinks to zero size. \\
The K\"ahler moduli space of these geometries presents a manifold richness of phenomena which provide
a natural testing ground for $A$-twisted topological string theory away from the large radius phase, 
as well as for the search of direct evidence for open/closed dualities in the strongly coupled $\alpha'$ regime. 
The pivotal example of the latter is given by the case $q=0$, 
which is the large $N$ dual background of the open $A$-model on $T^*L(p,1)$ obtained via geometric transition
\cite{Halmagyi:2003mm}. 
Moreover, it was noticed in \cite{Martelli:2004wu, Benvenuti:2004dy} that $\mathcal{X}_{p,q}$ can be ``blown-down'' 
to orbifolds of flat space of the form $\mathbb{C}^3/\mathbb{Z}_{p+q}$ (see table \ref{taborbif}); 
here the geometric picture becomes singular, though leaving still open the possibility for extracting enumerative 
results in terms of orbifold Gromov-Witten invariants. These are precisely the cases we will turn to study in section 
\ref{conti}.
\begin{table}[t]
\centering
\begin{tabular}{|c|c|c|c|}
\hline
$p$ & $q$ & $p+q$ & Weights \\
\hline
$1$ & $0$ & $1$ & $(0,0,0)$ \\
\hline
$1$ & $1$ & $2$ & $(0,1,1)$ \\
\hline
$2$ & $0$ & $2$ & $(0,1,1)$ \\
\hline
$2$ & $1$ & $3$ & $(1,1,1)$ \\
\hline
$2$ & $2$ & $4$ & $(1,1,2)$ \\
\hline
\end{tabular}
\caption{Orbifold degenerations of $\mathcal{X}_{p,q}$ into $\mathbb{C}^3/\mathbb{Z}_{p+q}$ for the first few values of $p$ and $q$. The fourth column lists the weights of the $\mathbb{Z}_{p+q}$ action on the coordinates $(z_1,z_2,z_3)$ of $\mathbb{C}^3$.}
\label{taborbif}
\end{table}

\section{Mirror symmetry for local CY and integrable systems}
\label{sectmirror}
\subsection{Period integrals} 
A procedure for constructing mirror duals of (among others) toric $CY$ threefolds has been provided in \cite{Hori:2000kt}, elaborating on previous results of \cite{Chiang:1999tz, Batyrev:1994hm}. The mirror geometry $\widehat{\mathcal{X}_{p,q}}$ of $\mathcal{X}_{p,q}$ is an affine hypersurface in $\mathbb{C}^2 \times (\mathbb{C}^*)^2 $ 

\beq
\label{curvamirrorpq}
x_1 x_2 =H_{p,q}(u,v)
\eeq
with $(x_1,x_2)\in \mathbb{C}^2$ and $(u,v)\in (\mathbb{C}^*)^2$.
In (\ref{curvamirrorpq}) $H_{p,q}(u,v)$ is the Newton polynomial \cite{Feng:2005gw} of the polytope 
$\Sigma_{p,q} \cap \{r_3=1\}$ in $\mathbb{Z}^3$ given by the intersection of the fan with 
the affine hyperplane $r_3=1$
\beq
\label{polynomialpq}
H_{p,q}(u,v) = a_1 v + \frac{a_2 u^{p-q}}{v} - \sum_{i=0}^{p}a_{i+3} u^i
\eeq
The geometry is therefore that of a quadric fibration over the $H_{p,q}(u,v)=\lambda \in \mathbb{C}$ plane, which degenerates to a node above the punctured Riemann surface $H_{p,q}(u,v)=0$. We will call the latter the \textit{mirror curve} $\Gamma_{p,q}$. \\
Now, mirror symmetry in the compact case prescribes to reconstruct the A-model prepotential from the computation of the (properly normalized\footnote{In this section, we will not be careful about normalization factors. This will be of course of our concern in the calculations of section \ref{conti}.}) periods of the holomorphic $(3,0)$ form $[\Omega] \in H^{3,0}(\widehat{\mathcal{X}_{p,q}})$ on a symplectic basis of homology three-cycles
$$\vec{\Pi}=\int_{\vec{\Gamma}\in H_3(\widehat{\mathcal{X}_{p,q}},\mathbb{Z})} \Omega$$
where $\Omega$ in this case would be the residue form on $H_{p,q}=0$ of the holomorphic $4$-form in $H^{4,0}(\mathbb{C}^2 \times (\mathbb{C}^*)^2 \setminus \widehat{\mathcal{X}_{p,q}})$
\beq
\label{omega}
\Omega = \hbox{Res}_{H_{p,q}(u,v)=x_1 x_2}\left[ \frac{dx_1 dx_2 du/u dv/v}{x_1 x_2-H_{p,q}(u,v)} \right]
\eeq
Special geometry then ensures \cite{Cox:2000vi} that the periods are related as

\beq
\vec{\Pi} = (t_0(a), t_i(a), \partial_{t_i} \mathcal{F}(a), 2\mathcal{F}-\sum_i t_i \partial_{t_i}\mathcal{F})
\eeq
where $t_i(a)$ defines a local isomorphism between the A model moduli space of $\mathcal{X}_{p,q}$ and the B model moduli space of $\widehat{\mathcal{X}_{p,q}}$, while $\mathcal{F}(t)$ is the prepotential, i.e. the sphere amplitude. They have respectively single and double logarithmic singularities at the large complex structure point \cite{Chiang:1999tz}. In the local case under scrutiny we must actually cope with the absence of a symplectic basis for $H_3(\widehat{\mathcal{X}_{p,q}},\mathbb{Z})$; according to \cite{Hosono:2004jp, Forbes:2005xt}, the formalism carries through to the local setting
by considering non-compact cycles as well and defining the periods along them via equivariant
localization.\footnote{This is really bothering solely for the case of local curves. For $p\geq 2$, as we will see, the integration over compact cycles is sufficient to extract enumerative information.} 
We will denote the corresponding extended homology as $H_3^{(ext)}(\widehat{\mathcal{X}_{p,q}},\mathbb{Z})$.
\\
The usual procedure the $(t_i, \partial_i\mathcal{F})$ are found is via integration of the associated $GKZ$ hypergeometric system \cite{Cox:2000vi, Chiang:1999tz}
\beq
\label{GKZeq}
\prod_{Q_i^{(k)}>0} \left(\frac{\partial}{\partial a_i}\right)^{Q_i^{(k)}} = \prod_{Q_i^{(k)}<0} \left(\frac{\partial}{\partial a_i}\right)^{Q_i^{(k)}} \qquad k=1,\dots, p
\eeq
with $Q_i^{(k)}$ as in (\ref{glsmpq}). In a patch of the B-model moduli space, these are the Picard-Fuchs ($PF$) equations for $\widehat{\mathcal{X}_{p,q}}$. Typically, solutions of (\ref{GKZeq}) are obtained via series integration by Frobenius method, i.e. solving recursion relations for the coefficients of a series expansions for $(t_i, \partial_i\mathcal{F})$. This has to be done patch by patch in the moduli space though, and it turns out to be hardly practicable as soon as the number of K\"ahler classes increases. To our knowledge, there is no explicit solutions in the literature when $h_{11}(\mathcal{X}_{p,q})>2$. \\
A possible alternative way to find $\vec{\Pi}$ is via direct integration. This not viable in the general case, but notice that due to the particular form (\ref{curvamirrorpq}) of $\widehat{\mathcal{X}_{p,q}}$, the integration of $\Omega$ over three-cycles boils down to that of a $1$-differential $d\lambda$ over a basis of cycles in $H_1^{(ext)}(\Gamma_{p,q},\mathbb{Z})$. 
As shown in \cite{Forbes:2005xt}, the periods of $\Omega$ solve the $PF$ system (\ref{GKZeq}) if and only if those of
\beq
d\lambda_{p,q} \equiv \hbox{Res}_{H_{p,q}(u,v)=0} \left[\frac{du dv}{uv}\log{H_{p,q}(u,v)}\right]
\eeq
do on a basis of $H_1^{(ext)}(\Gamma_{p,q},\mathbb{Z})$. Picking up the residue gives

\beq
\label{differpq}
d\lambda = \log{v} \frac{du}{u}
\eeq
and the periods are thus computed as
\beq
\Pi^{\Gamma_{p,q}}_\gamma = \int_{\gamma \in H_1^{(ext)}(\Gamma_{p,q}, \mathbb{Z})} \log{v} \frac{du}{u}
\eeq
Unfortunately, the integrals are typically too awkward to carry out and no expression is known except for the simplest case of local curves; a perturbative evaluation of them, though clearly possible, has no real advantages compared to tackling the $PF$ system upfront. However, we will see in section \ref{conti} how to handle them in a direct way.\\
\subsubsection{Mirror symmetry for open strings}
\label{openstrings}
\begin{figure}[b]
\centering
\includegraphics{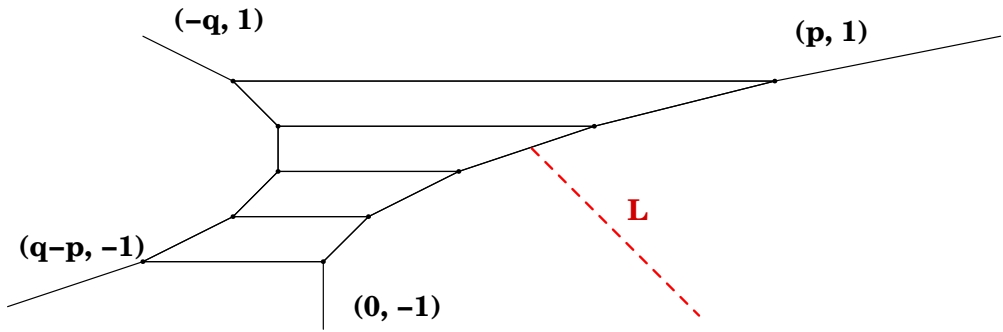}
\caption{The pq-web for $p=5$, $q=2$ with a lagrangian brane on an inner leg.}
\end{figure}
Recently, the open string sector of the $A$-model on toric $CY$ has been subject to deeper investigation, following the insight of \cite{Aganagic:2000gs, Aganagic:2001nx}, where a class of special Lagrangian submanifolds was constructed generalizing \cite{Harvey:1982xk}. The prescription of \cite{Aganagic:2000gs} relies on the realization of a toric $CY$ as a (degenerate) $\mathbb{T}^3$ fibration, parameterized by  $\theta_i$, over the $|z_i|$ base, see (\ref{dtermglsm}, \ref{gfixglsm}). 
The authors of \cite{Aganagic:2000gs} consider a $3-k$ real dimensional linear subspace $W$
of the base 
\beq
\label{dtermbrana}
\sum_i q_i^\alpha |z_i|^2 = c^\alpha \qquad \alpha=1,\dots, k ; \quad q_i^\alpha \in \mathbb{Q}
\eeq
and then specify a $\mathbb{T}^k$ fibration $L$ over this subspace in such a way that 
the K\"ahler form $\omega = \sum_i d|z_i|^2 \wedge d\theta_i$ vanishes on it 
\beq
\omega|_L=0 
\label{gfixbrana}
\eeq
The total space of this fiber bundle $L$ is then Lagrangian by construction; moreover, it turns out that it is 
volume minimizing in its homology class ({\it special Lagrangian}) if and only if $\sum_i q_i^\alpha =0$. In
this case (\ref{gfixbrana}) implies 
\beq
\sum_i \theta_i =0
\label{susy}
\eeq
In the case in which $c^\alpha$ in (\ref{dtermbrana}) are such that $W$ intersects the edges of the toric web, i.e. the loci where one $S^1$ of the toric fibration shrinks, $L$ splits into two Lagrangians $L_\pm$ with topology $\mathbb{R}^2 \times S^1$: the open modulus $z$ is then given by the size of the circle, complexified with the holonomy of 
a $U(1)$ connection along it. \\
The mirror symmetry construction of \cite{Hori:2000kt} has been extended to these brane
configurations in \cite{Aganagic:2000gs}. When $k=2$, $L_+$ (resp. $L_-$) gets mirror mapped to a curve parameterized by 
$x_2$ (resp. $x_1$)
\beq
H_{p,q}(u,v)=0=x_1 \quad (\hbox{resp.} =x_2)
\eeq
The moduli space of the mirror brane is then simply the mirror curve $\Gamma_{p,q}$. Picking a parametrization thereof (for instance the projection on the $u$ or $v$-lines) by a complex variable $z$ leads us to write the open topological partition function
\beq
\mathcal{F}_{open}(\{t_i\},z)= \sum_{g,h} \mathcal{F}_{g,h}(\{t_i\}) z^h
\eeq
 the sum is both over the genus of the source curve and the number of connected components of its boundary. The choice of a ``good'' parametrization is dictated by mirror symmetry and is related to phase transitions in the open string moduli space for branes ending on toric curves meeting at a vertex in the web (see \cite{Bouchard:2007ys} for details). \\
A very important fact is that the meromorphic differential $d\lambda$ turns out to have a significant role for open string amplitudes as well. The dimensional reduction of the holomorphic Chern-Simons action on the mirror brane indeed yields a particularly simple expression for the disc amplitude $g=0$, $h=1$. It is simply given by the ``Abel-Jacobi'' map 

\beq
\label{abeljac}
\mathcal{F}_{0,1}(t,z)=\int^z \log{v(u)} \frac{du}{u} 
\eeq
where the integral on the r.h.s. is a chain integral $[z^*,z]$, with $z^*$ a fixed point on the mirror curve. 
In \cite{ov} it was noticed that the disc amplitude (\ref{abeljac})
in a suitable parameterization gets the form 
\beq
\mathcal{F}_{0,1}(t,z_{open}) = \sum_{m,n} N_{m,n} Li_2(e^{-t\cdot m} z_{open}^n)
\label{discflat}
\eeq
In (\ref{discflat}) 
$N_{m,n}$ are integer numbers counting open string BPS states
and $z_{open}$ is the dressed open coordinate \cite{Aganagic:2001nx}
\beq
\label{openflat}
z_{open} = z + \sum_i \frac{w_i-t_i}{r_i}
\eeq
where $w_i$ are combinations of gauge-invariant sigma model variables vanishing at the point of maximally unipotent monodromy

\beq
w_i = \prod_j a_j^{Q_j^{(i)}} = t_i + \mathcal{O}(e^{-t_i})
\eeq
and $r_i$ are rational numbers. Notice that the open flat coordinates gets corrected by 
\textit{closed} worldsheet instantons only. As discovered in \cite{Lerche:2001cw},(see also \cite{Forbes:2003ki}) an extended Picard-Fuchs system may be constructed such that (\ref{abeljac}), (\ref{openflat}) are in its kernel and this can be used for determining the $r_i$ in (\ref{openflat}).

\subsection{The $B$-model moduli space}
\label{Bmoduli}
A few remarks are in order at this point. The Riemann surfaces $\Gamma_{p,q}$ come in a family parameterized by  
$\{a_i\}$ in (\ref{curvamirrorpq}), which are the complex moduli of the mirror geometry.
The curve $\Gamma_{p,q}$ will be generically smooth in the $B$-model moduli space: 
we will denote the open set where this happens as $\mathcal{M}^B_{p,q}$. However, a compactification of 
$\mathcal{M}^B_{p,q}$ will lead to loci where this is no longer true. Indeed, $\Gamma_{p,q}$ degenerates to 
a singular curve on the so-called principal discriminant locus of the $PF$ system (\ref{GKZeq}). In 
correspondence with this, one of the homology cycles of $\Gamma_{p,q}$ shrinks to zero size; $GKZ$ solutions have 
then singularities and are subject to logarithmic monodromy transformations around these loci, 
to which we will refer to as \textit{conifold} loci. 
Moreover, there also regions in the $B$-model moduli space where the curve $\Gamma_{p,q}$ stays smooth, but the periods have
\textit{finite} monodromy because the moduli space itself is singular, locally looking like $\mathbb{C}^p/\mathbb{Z}_n$; 
at the conformal field theory level, this would be reflected by the appearance of a discrete quantum symmetry. 
We will refer to the latter as an {\it orbifold} phase, in which we still retain a geometric picture, 
though involving a singular (orbifold) target space in the $A$-model side. \\
To see why this happens from the mirror perspective one might argue as follows. The $a_i$'s are 
sort of homogeneous coordinates for the complex structure moduli space of $\widehat{\mathcal{X}^{p,q}}$. 
Indeed, only $p$ out of $p+3$ are really independent, as an overall rescaling of them and scalings of $u$ and $v$ in 
(\ref{curvamirrorpq}) leave invariant the symplectic form 
\beq
\frac{du}{u}\wedge\frac{dv}{v}
\eeq
in $(\mathbb{C}^*)^2$. That is, the moduli space of the mirror theory might be seen as arising from a holomorphic quotient 
of $\mathbb{C}^{p+3}$ by a $(\mathbb{C}^*)^3$ action with charges (see (\ref{polynomialpq}))

\beq
\label{secondaryglsm}
\begin{array}{c|cccccccc}
&  a_1 & a_3 & a_{p+3} & a_2 & a_4 & \dots & a_{p+2} \\
\hline
Q_1 & 1,& 0,& 0,& -1,& 0, &  \dots, & 0 \\
Q_2 & 0,& 0, & p, & p-q, & 1, & \dots, & p-1 \\
Q_3 & 1, & 1, & 1, & 1, & 1, & \dots, & 1 \\
\end{array}
\eeq
By subtracting a suitable codimension $>0$ locus to $\mathbb{C}^{p+3}$, we thus end up with 
a \textit{toric} compactification of the family $\mathcal{M}^B_{p,q}$, which we call $\overline{\mathcal{M}^{B,tor}_{p,q}}$. 
Remarkably enough, inspection shows \cite{Cox:2000vi} that the skeleton of a fan for the above system of charges 
is simply given by the columns of the $GLSM$ (\ref{glsmpq}), and the toric variety associated with it is complete.
This fan is called the {\it secondary fan} of $\mathcal{X}^{p,q}$. In fact, strictly speaking we are not dealing with a toric variety, as typically the secondary fan will contain non-smooth simplicial cones, perhaps with marked points along their facets. In the latter case, this would mean that the patch parameterized by the corresponding $a_i$'s looks like $\mathbb{C}^p/\mathbb{Z}_n$ rather than $\mathbb{C}^p$; as such, the periods of the holomorphic three-form will inherit the finite monodromy from the monodromy of the $a_i$ themselves. This will be of fundamental importance in our study of the $[\mathbb{C}^3/\mathbb{Z}_4]$ orbifold in section \ref{conti}. \\
Additional, but somewhat milder phase transitions involve the purely open string sector as well and are related, as already anticipated, to a choice of a parametrization of $\Gamma_{p,q}$. See \cite{Bouchard:2007ys}, to which we refer for a complete discussion of this subject.

\subsection{Relation with integrable systems and five-dimensional gauge theories}
\label{legametoda}
Interestingly, the mirror curves of $\mathcal{X}_{p,q}$ geometries are related to the Seiberg-Witten curve
of five dimensional gauge theories and the related integrable systems. Since, as we will show in the next section, this observation
will prove to be very fruitful in the study of the topological string moduli space of $\mathcal{X}_{p,q}$, we describe it here in some detail. \\
First of all, let us rewrite (\ref{polynomialpq}) as 

\beq
\label{curvamirrorpq2}
Y^2 = P_p(X)^2 - 4 a_1 a_2 X^{p-q}
\eeq
upon setting

$$Y=a_1 v - a_2 u ^{p-q}/v$$ 
$$ X=u$$
\beq P_p(X)=\sum_{i=0}^p a_{i+3} X^i\label{Pp}\eeq
In the $a_1=a_2=(\Lambda R)^p$, $a_3=a_p=1$ patch the curve (\ref{curvamirrorpq2}) and the differential (\ref{differpq})
are precisely the Seiberg-Witten curve and differential of $SU(p)$ ${\cal N}=1$ SYM theory on $\mathbb{R}^4\times S^1$ with a $q$-dependent Chern-Simons term \cite{Hollowood:2003cv, Tachikawa:2004ur}. \\

Moreover, the SW curve and differential in the case $p=q$ were shown in \cite{Nekrasov:1996cz} to coincide with the spectral curve and action differential of the $A_{p-1}$ Ruijsenaars model \cite{ruijsenaars}, i.e. the $A_{p-1}$ periodic relativistic Toda chain. More precisely, setting $\zeta=\Lambda R$, (\ref{polynomialpq}) reads for $p=q$

$$\Gamma_{p,p}: \quad \zeta^p (v+\frac{1}{v})= 1+\sum_{l=1}^{p-1} u^l S_l + u^p, \quad d\lambda_{p,p}=\log{v }  \frac{du}{u}$$
which can be rewritten as
\beq
\label{curvatodarel}
\det(L(z)-w) = \sum_{j=0}^{p}(-w)^{p-j}\sigma_j(z) = 0
\eeq
with the Lax matrix defined as

\bea
L_{ij} & =& e^{Rp_{i}} f _{i} (l_{ij} + b_{ij})\nonumber \\
l_{ij}  &=&  \delta_{i,j+1} (1 + \zeta^{p} z) \xi_{i} -
\delta_{i,1}\delta_{j,p} (1 + \zeta^{-p} z^{-1}) \xi_{1} \nonumber \\
b_{ij}  &=&   \Biggl[
\begin{array}{cc}
-(i\zeta)^{p} & \quad i \leq j-1 \\
1 &\quad i > j-1
\end{array} \nonumber \\
f_{i}^{2}  & =& (1 - \zeta^{2} e^{q_{i+1}-q_{i}} )(1 -  
\zeta^{2}
e^{q_{i}-q_{i-1}} )\nonumber \\
\xi_{i}^{-1} & =& 1 - \zeta^{-2} e^{q_{i-1} - q_{i}}
\label{lax}
\eea
where $q_{p+1} = q_{1}, q_{0} = q_{p}$, $\sigma_j$ are the elementary symmetric functions of $L(z)$, $S_j$ their $z$-independent 
factor,  and we have made the change of variables \cite{Nekrasov:1996cz}
$$-w u = 1+\zeta^p z, \quad z=v$$
An identification of the curves for $q<p$ as the spectral curves of some finite dimensional integrable mechanical system seems to be presently not known, and it would be interesting to understand to role of the $q$ parameter in this context. \\

A second important remark is about the ``field theory limit'' discussed in \cite{Katz:1996fh}. From the mechanical system point of view, the parameter $\zeta$ in (\ref{lax}) is essentially the inverse of the speed of light, while in the field theory perspective $\zeta = \Lambda R$, where $\Lambda$ is the strong coupling scale and $R$ is the radius of the fifth-dimensional circle. This means that the four-dimensional limit might be achieved as the non-relativistic limit of the Toda chain. Denoting with $e^{\phi_i}$, $i=1\dots,p$  the roots of the polynomial $P_p(x)$ (\ref{Pp}) and introducing the new set of variables \cite{Hollowood:2003cv}

\bea
y &=& Y \nonumber \\
X &=& e^{2R x} \\
\label{xpiccolo}
e^{\phi_i} &=& e^{R (a_i-a_{i+1})} \nonumber 
\eea
we have that in the $R\to 0$ limit (\ref{curvamirrorpq2}) reduces to

\beq
\label{SW4d}
y^2 = \tilde P_p(x)^2-4 \Lambda^{2p}
\eeq
which is the Seiberg-Witten curve of $\mathcal{N}=2$ $SU(p)$ Super Yang-Mills in $d=4$. Notice that in the $R\to 0$ limit we completely lose track of the Chern-Simons parameter $q$, which has disappeared in formula (\ref{SW4d}). More importantly, the variable $x$ in (\ref{xpiccolo}) takes now values in $\mathbb{C}$ in contrast with the $\mathbb{C}^*-$variable $X$. Because of this, the peculiarly five-dimensional form of the differential (\ref{differpq}), i.e. $d\lambda = \log{v} d \log{u}$, is replaced in the $R\to 0$ limit by that of the usual (non-relativistic) Toda differential $d\lambda= x d\ln(\tilde{P_p}+y)$. In fact, as we will see in the following, the relativistic system and its non-relativistic counterpart - which by the discussion above coincide respectively with the $A-$model on $\mathcal{X}_{p,q}$ and with its $4d$ Seiberg-Witten limit - bear still deep structural resemblances and our aim will be to try to exploit this to our advantage.


\section{Solving the $GKZ$ system in the full moduli space}
\label{conti}
In this section we provide a method for finding the mirror map, as well as the sphere and disc amplitude, for the $A$-model on $\mathcal{X}_{p,q}$ to all orders in $\alpha'$ without resorting to solving the $GKZ$ system directly. This will be accomplished by finding closed forms for derivatives of the period integrals $\Pi^{\Gamma_{p,q}}_\gamma$ w.r.t. the bare moduli as generalized hypergeometric functions. \\

First of all, let us resume what the ingredients at our disposal are. According to (\ref{curvamirrorpq2}) 
the mirror curve $\Gamma_{p,q}$ is a two-fold covering of the $X$ plane branched at $Y(X)=0$, that is the locus

\beq
\label{eqbranchpq}
P_p(X)^2 = 4 a_1 a_2 X^{p-q}
\eeq
The resulting curve has genus $p-1$ and four punctures corresponding to the two inverse images of $X=0$, $X=\infty$. Let us denote the solutions to (\ref{eqbranchpq}) as $\{b_i\}_{i=1}^{2p}$.  A basis for $H_1^{(ext)}(\Gamma_{p,q},\mathbb{Z})$ might be taken as the circles $A_i$, $B_i$ encircling the intervals
\beq
I_{A_i}=[b_{2i-1}, b_{2i}] \qquad I_{B_i}= [b_{2i}, b_{2i+1}]
\eeq
for $i=1, \dots, p-1$, plus a circle $A_0$ around one of the punctures at $X=0$ and a contour $B_0$ connecting the two punctures at $X=0$ and $X=\infty$. 
\begin{figure}[t]
\centering
\includegraphics[scale=0.5]{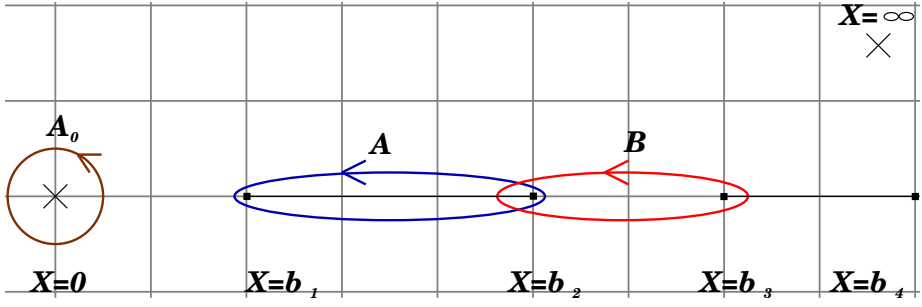}
\caption{Cuts and punctures of the $X$ plane in the genus $1$ case.}
\label{tagli}
\end{figure}
The 1-differential $d\lambda_{p,q}$ is given, in  an affine patch parameterized by $X$, as

\beq
\label{differpq2}
d\lambda_{p,q}(X)= \log v(u) \frac{du}{u} 
=\log\left(\frac{P_p(X) \pm \sqrt{P_p(X)^2-4 a_1 a_2 X^{p-q}}}{2 a_1} \right)\frac{dX}{X}
\eeq
and a complete set of periods can be obtained by integrating it over the $A/B$-cycles
\beq
\Pi_{A/B} = \oint_{A/B} d\lambda_{p,q}
\label{boh}
\eeq
More explicitly,
\bea
\Pi_{A_i}& = &\int_{b_{2i-1}}^{b_{2i}} \log\left(\frac{P_p(X) +  \sqrt{P_p(X)^2-4 a_1 a_2 X^{p-q}}}{P_p(X) -  \sqrt{P_p(X)^2-4 a_1 a_2 X^{p-q}}} \right) \frac{dX}{X}  \\
\Pi_{B_i}& = &\int_{b_{2i}}^{b_{2i+1}} \log\left(\frac{P_p(X) + \sqrt{P_p(X)^2-4 a_1 a_2 X^{p-q}}}{P_p(X) - \sqrt{P_p(X)^2-4 a_1 a_2 X^{p-q}}} \right) \frac{dX}{X} \\
\label{residuopq}
\Pi_{A_0} &=& \oint_{X=0} \log\left(\frac{P_p(X) \pm \sqrt{P_p(X)^2-4 a_1 a_2 X^{p-q}}}{2 a_1} \right)\frac{dX}{X} \\
\Pi_{B_0} &=& \int_{0}^\infty \log\left(\frac{P_p(X) \pm \sqrt{P_p(X)^2-4 a_1 a_2 X^{p-q}}}{2 a_1} \right)\frac{dX}{X}
\eea
We now make the following observation. 
As we have already noticed, the curve (\ref{curvamirrorpq2}) and the differential (\ref{differpq2}) are the Seiberg-Witten (SW) curve and differential of a five dimensional theory compactified on a circle. In Seiberg-Witten theory, the gauge coupling matrix  
\beq
\tau_{ij}=\frac{\partial \Pi_{B_i}}{\partial u_k} \left(\frac{\partial \Pi_{A_k}}{\partial u_j}\right)^{-1}
\eeq
where $u_i$ are Weyl-invariant functions of the scalar fields, is known to be the period matrix of the {\it compactified} SW curve,
that is a ratio of periods of holomorphic differentials. 
We then expect that derivatives of $d\lambda_{p,q}$ with respect to suitable functions of the bare moduli are \textit{holomorphic differentials} on the compactified $\overline\Gamma_{p,q}$

\beq
[\partial_{f(a_i)}d\lambda] \in H^{1,0}(\overline\Gamma_{p,q})
\eeq
This is substantiated by the fact that, for $p=q=2$, the relativistic Toda system and the non-relativistic one share the same oscillation periods \cite{degaruij}; more precisely, the derivatives of the action with respect to the energy are the same (elliptic) functions of the bare parameters. This was also noticed in \cite{Nekrasov:1996cz} in the study of the singularities of the moduli space of $\mathcal{N}=1$ $SU(2)$ $SYM$ in $d=5$. \\
Explicitly, we indeed have

\beq
\label{holdiffpq}
\frac{\partial d\lambda_{p,q}}{\partial a_{j+4}} = \frac{X^{j}}{\sqrt{P^2_p(X)-4 a_1 a_2 X^{p-q}}}dX
\eeq
i.e., for $j=0,\dots, p-2$, a basis of holomorphic $1$-forms on the $4$-point compactification $\overline{\Gamma}_{p,q} = \Gamma_{p,q} \cup \{0_+, 0_-, \infty_+, \infty_- \}$ of the spectral curve $\Gamma_{p,q}$.

\subsection{Period integrals and Lauricella functions}
\label{sezlauric}
This last observation allows us to give a straightforward recipe for computing series expansions of solutions of the $GKZ$ system (\ref{GKZeq}) in the full $B$-model moduli space. The procedure is the following: 
\begin{enumerate}
\item start with $\Pi_{A_i/B_i}$ and consider its $a_{j+4}$ derivative for $0\leq j\leq p-2$
\beq
\label{hyperintpq}
\frac{\partial \Pi_{A_i/B_i}}{\partial a_{j+4}} = \int_{e_i}^{e_{i+1}} \frac{X^j}{\sqrt{\prod_{i=1}^{2p} (X - b_i)}} dX
\eeq
with $e_i=b_{2i-1}$, $e_i=b_{2i}$ for the $A$ and the $B$ cycles respectively. The hyperelliptic integral (\ref{hyperintpq}) has a closed expression given in terms of multivariate generalized hypergeometric functions of Lauricella type \cite{extonfun}
\bea
\label{derperiodpq}
\frac{\partial \Pi_{A_i/B_i}}{\partial a_{j+4}} &=& 
{\rm e}^{i \varphi} \pi \frac{(e_i)^j}{\sqrt{\prod_{k\neq i, i+1} (e_k-e_i)}} \nonumber \\
& \times &  F_D^{(2p-1)} \left(\frac{1}{2}; \frac{1}{2}, \dots, \frac{1}{2}, j; 1; x_1, \dots, \widehat{x_i}, \widehat{x_{i+1}}, \dots, x_{2p}, \frac{e_{i+1}-e_i}{e_i}\right) \nonumber \\
\eea
where $x_j=(e_{i+1}-e_i)/(e_j-e_i)$, $2 \varphi= l \pi$, $l\in \mathbb{Z}$ is a phase depending on $x_i$ and $F_D^{(n)}$ is the hypergeometric series
\beq
F_D^{(n)}(\alpha;\{\beta_i\};\gamma; \{\delta_i\}) = \sum_{m_1 \dots m_n=0}^{\infty} 
\frac{(\alpha)_{m_+\dots+m_n}(\beta_1)_{m_1}\dots (\beta_n)_{m_n} \delta_1^{m_1}\dots \delta_n^{m_n}}{(\gamma)_{m_+\dots+m_n}
m_1!\dots m_n!}
\eeq
which converges when $|\delta_i|<1$ for every $i$. In the above formula we used the standard Pochhammer symbol
$(\alpha)_m = \Gamma (\alpha + m)/\Gamma(\alpha)$.
There are many alternative ways to express (\ref{hyperintpq}), for instance in terms of hyperelliptic $\theta$ functions; 
however,  
the above expression proves to be useful due to the fact that Lauricella $F_D^{(n)}$ has good analytic continuation properties outside the unit polydisc $|\delta_i|<1$; some formulae, as well as asymptotic expansions around singular submanifolds, are collected in the Appendix, while others can be found in \cite{extonfun, extonint}. Notice that, as opposed to the usual situation in solving $PF$ equations by Frobenius method, we are not dealing here with hypergeometric functions of the bare moduli, but rather of the relative distance $x_i$ between ramification points; they have singular values precisely when the latter becomes $0$, $1$ or infinity, that is when we encounter a pinching point of $\Gamma_{p,q}$. This shift in perspective is definitely an advantage compared to other expressions for hyperelliptic integrals, involving for instance the $F_4$ Appell function for genus 2 \cite{Klemm:1995wp, Ohta:1998yu}. These are simpler functions of the bare moduli, but have worse analytic continuation properties and are less suited for a more complete study of the moduli space, regarding for instance intersecting submanifolds of the principal discriminant locus. The above fact was already pointed out in \cite{Akerblom:2004cg}, where the properties of $F_D^{n}$ were exploited to study the $\mathbb{Z}_3$ point of $\mathcal{N}=2$ $SU(3$) $SYM$. \\
In many cases, $F_D^{(n)}$ can be reduced to a more familiar form. For instance, for $p=2$ we have the expected complete elliptic integrals of the first kind

\bea
\label{derperiodA2}
\frac{\partial \Pi_{A}}{\partial a_4} &=& \frac{2}{\sqrt{(b_1-b_3)(b_2-b_4)}}K\left[\frac{(b_1-b_2)(b_3-b_4)}{(b_1-b_3)(b_2-b_4)}\right] \\
\label{derperiodB2}
\frac{\partial \Pi_{B}}{\partial a_4} &=& \frac{2}{\sqrt{(b_1-b_2)(b_3-b_4)}}K\left[\frac{(b_1-b_3)(b_2-b_4)}{(b_1-b_2)(b_3-b_4)}\right]
\eea
\item Once we have a representation for the derivatives of the periods in the form (\ref{derperiodpq}), (\ref{derperiodA2})-(\ref{derperiodB2}) we can use the formulae in Appendix \ref{applauric} to analytically continue them in any given patch of the $B$-model moduli space and find a corresponding power series expansion in the bare moduli $a_i$. Integrating back with respect to $a_j$ yields $\Pi_{A_i}$ and $\Pi_{B_i}$ up to a constant of integration, independent of $a_j$ for $0\leq j\leq p-2$. This has to be fixed either by some indirect consideration (for instance, by imposing a prescribed asymptotic behavior around a singular point) or by plugging it inside the $PF$ system and imposing that the period be in the kernel of the $GKZ$ operators. This operation leads to a closed $ODE$ integrable by quadratures, which completes the solution of the problem of finding expansions for $\Pi_{A_i/B_i}$ everywhere in the $B$-model moduli space.
\item The procedure provides us with $p-1$ flat coordinates as well as $p-1$ conjugate periods out of which to extract the prepotential. In order to find the $p^{th}$ modulus, we pick up the residue (\ref{residuopq})

\beq
\oint_{X=0_\pm} d\lambda = \left\{
\begin{array}{cc}
\log\left(\pm \frac{a_3}{a_1}\right) & \hbox{for } q<p \\
\log\left(\frac{a_3\pm\sqrt{a_3^2-4 a_1a_2}}{2a_1}\right) & \hbox{for } q=p
\end{array}\right.
\label{residuolog}
\eeq
which are manifestly solutions of (\ref{GKZeq}). In the following, we will choose an appropriate combination of them in order to have a prescribed behavior around the expansion point under scrutiny.

\item Closed form computation of derivatives with respect to $a_j$ can be done for open string amplitudes as well, which might be used to trade an expansion in terms of the $z$ parameter in (\ref{openflat}) with one in $a_j$, completely resummed w.r.t. $z$. In this case, dealing with chain integrals instead of period integrals leads one to consider indefinite integrals and thus incomplete hyperelliptic integrals. The latter can still be given the form of a multivariate Lauricella function, but with order increased by one \cite{extonfun}

\bea
\label{incompletelauricella}
\frac{\partial \mathcal{F}_{0,1}(\{a_k\},z)}{\partial a_{j+4}} &=& 
{\rm e}^{i \varphi} \pi \frac{(e_i)^j}{\sqrt{\prod_{k\neq i, i+1} (e_k-e_i)}}
2\sqrt{z} \\
& \times & F_D^{(2p)} \left(\frac{1}{2}; \frac{1}{2}, \dots, \frac{1}{2}, j, \frac{1}{2};\frac{3}{2}; x_3, \dots, x_{2p}, \frac{e_{2}-e_1}{e_1}, z\right) \nonumber
\eea
As before, for $p=2$ (\ref{incompletelauricella}) boils down to an incomplete elliptic integral of the first kind in the form

\beq
\label{incompleteelliptic}
\frac{\partial \mathcal{F}_{0,1}(\{a_k\},z)}{\partial a_{4}} = 2  \sqrt{\frac{1}{\tilde{b}}} F\left(\sin
   ^{-1}\left(\sqrt{\frac{(b_{1}-b_{4})
   (b_{2}-z)}{(b_{2}-b_{4})
   (b_{1}-z)}}\right) \Bigg|\frac{\tilde{a}}{\tilde{b}}\right)
\eeq
where
$$
\tilde{a}=(b_2-b_4)(b_1-b_3) \qquad \tilde{b}=(b_2-b_3)(b_1-b_4)
$$
\end{enumerate}

\vspace{.3cm}

Another important advantage of this method is that,
instead of integrating back patch-wise with respect to $a_j$, 
we can get our hands dirty and work directly with an Euler-type integral representation of the periods. 
The fact that $F_D^{(n)}$ has a single integral representation saves us most of the pain 
in the problem of finding the explicit analytic continuation of $\Pi_{A_i/B_i}$, 
which in the multi-parameter case involves the use of multi-loop Mellin-Barnes integrals.
The details for the case $p=q=2$ which will be of our interest later on for the computation of orbifold Gromov-Witten invariants
are reported in Appendix \ref{analytic}, where also a closed expression for the A-period can be found in terms of
a generalized Kamp\'e de F\'eriet hypergeometric function.

\section{Warm-up tests of the formalism}
Let us show how the steps described in Section \ref{sezlauric} allow to quickly recover some known results about mirror symmetry for local surfaces.

\begin{figure}[!b]
\centering
\includegraphics{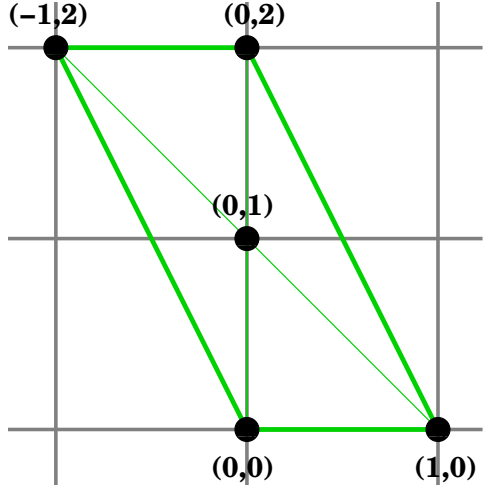}
\caption{The fan of local $\mathbb{F}_0$.}
\end{figure}

\subsection{Local $\mathbb{F}_0$ : mirror map at large radius}
Local mirror symmetry for $K_{\mathbb{F}_0}$  has been studied in \cite{Aganagic:2002wv} in the check of the large $N$ duality with Chern-Simons theory on $S^3/\mathbb{Z}_2$. The mirror curve in this case can be written as
\beq
\label{curvaf0}
a_1 v + a_2/v= a_3/u + a_4 + a_5 u
\eeq
Good variables around the large complex structure point \cite{Chiang:1999tz} are given by
\beq
\label{zBzF}
z_B=\frac{a_1 a_2}{a_4^2} \qquad z_F=\frac{a_3 a_5}{a_4^2}
\eeq
Let us use the scaling freedom (\ref{secondaryglsm}) to set 
\beq
a_3=a_5=1, \qquad a_1=a_2
\eeq
By using the change of variables (\ref{curvamirrorpq2}) the curve (\ref{curvaf0}) is then given by
\beq
\label{curvaf0lr}
Y^2=\left(X^2+\frac{X}{\sqrt{z_F}} + 1\right)^2-\frac{4z_B}{z_F} X^2
\eeq
which is a double covering of the $X-$plane branched at
\bea
\label{branchf0lr}
b_1 &=& \frac{-1+2\sqrt{z_B}-\sqrt{1-4\sqrt{z_B}+4z_B-4z_F}}{2\sqrt{z_F}} \\
b_2 &=& \frac{-1-2\sqrt{z_B}-\sqrt{1-4\sqrt{z_B}+4z_B-4z_F}}{2\sqrt{z_F}} \\
b_3 &=& \frac{-1+2\sqrt{z_B}+\sqrt{1-4\sqrt{z_B}+4z_B-4z_F}}{2\sqrt{z_F}} \\
b_4 &=& \frac{-1-2\sqrt{z_B}+\sqrt{1-4\sqrt{z_B}+4z_B-4z_F}}{2\sqrt{z_F}}
\eea
We choose the $A-$cycle as the loop encircling $[b_1, b_2]$. 
The asymptotics of the corresponding period will indeed identify it as the flat coordinate around $z_B=z_F=0$. 
By expanding (\ref{derperiodA2}) in $(z_B,z_F)$ we have
\bea
\frac{\partial \Pi_A}{\partial a_4} &=& \sqrt{z_F} (20 z_B^3+6 (30 z_F+1) z_B^2\nonumber \\ &+& 2 (90 z_F^2+12 z_F+1) z_B+20 z_F^3+6 z_F^2+2
   z_F+1) + \dots 
\eea
which integrates to
\beq
\label{PiAF0}
\Pi_A = \log (z_F)+\frac{20 z_B^3}{3}+60 z_F z_B^2+3 z_B^2+60 z_F^2 z_B 
\eeq
From (\ref{residuolog}) and (\ref{zBzF}) we can compute the remaining flat coordinate as
\beq
\label{Pi0F0}
\Pi_0 = -\frac{1}{2} \log \frac{z_B}{z_F}
\eeq
It is then easy to see that the combinations of periods that has the
right asymptotics at large radius are given by
\beq
-t_B\equiv -2 \Pi_0(z_B,z_F)+\Pi_A(z_B,z_F), \qquad -t_F\equiv\Pi_A(z_B,z_F)
\eeq
Inversion of (\ref{PiAF0}) and (\ref{Pi0F0}) reads, setting $Q_B=e^{-t_B}$, $Q_F=e^{-t_F}$
\bea
z_B &=& 6 Q_B^3-2 Q_B^2+6 Q_F^2 Q_B-2 Q_F Q_B+Q_B+\dots \nonumber \\
z_F &=& 6 Q_F^3-2 Q_F^2+6 Q_B^2 Q_F-2 Q_B Q_F+Q_F+\dots
\eea
which is the mirror map as written in \cite{Marino:2006hs}. 

\subsection{Local $\mathbb{F}_0$ : orbifold point} 

Analogously, we can write down the expansion for the orbifold point \cite{Aganagic:2002wv}, which corresponds to 
$a_1=a_2=a_3=a_5=1$, $a_4=0$. Setting $a_1=a_2=\sqrt{1-x1}$, $a_4=x_1 x_2$, $a_3=a_5=1$ as in \cite{Aganagic:2002wv}, we have
\bea
s_1 & \equiv & \Pi_0=-\log(1-x_1) \nonumber \\
s_2 & \equiv & \Pi_A + \Pi_B/2 = 
\frac{1}{61931520 \pi }\bigg[x_2 (35 (32 (x_1-2) x_1 (x_1 (11 x_1-96)+96) E(x_1)\nonumber \\
&+& x_1 (x_1 (x_1 (x_1 (105
   x_1-1856)+8000)-12288)+6144) K(x_1)) x_2^8 +\dots \bigg] \nonumber
\eea
Upon introducing $\tilde s_1=s_1$ and $\tilde s_2=s_1/s_2$ we have
\bea
x_1(\tilde s_1) &=& 1-e^{-\tilde s_1} \nonumber \\
x_2(\tilde s_1, \tilde s_2) &=& 
\tilde s_2+\frac{\tilde s_2}{4}  \tilde s_1+\bigg(\frac{\tilde s_2}{192}-\frac{\tilde s_2^3}{192}\bigg)
   \tilde s_1^2+\bigg(-\frac{\tilde s_2}{256}-\frac{\tilde s_2^3}{768}\bigg) \tilde s_1^3+\bigg(-\frac{49 \tilde s_2}{737280} \nonumber \\ &+& \frac{7
   \tilde s_2^3}{73728}-\frac{7 \tilde s_2^5}{245760}\bigg) \tilde s_1^4+\bigg(\frac{5 \tilde s_2^3}{98304}-\frac{7
   \tilde s_2^5}{983040}\bigg) \tilde s_1^5 + \dots
\eea
in perfect agreement with \cite{Aganagic:2002wv}. Needless to say, the prepotential computation can be checked exactly the same way. We have
\bea
F_{s_2} \equiv \Pi_A  &=& 
\frac{1}{53760}\bigg[x_1 x_2 \Big((75 x_1^3 x_2^6-56 x_1^2 (10 x_2^2+9) x_2^4+64 x_1 (10 x_2^4 +21
   x_2^2 \nonumber \\ &+& 70) x_2^2 
- 107520\Big) K(1-x_1)+\dots\bigg]
\nonumber \\
&=& \log\left(\frac{x_1}{16}\right) s_2 
-\frac{x_2^3}{12} x_1+(\frac{x_2}{4} +\frac{x_2^3}{48}) x_1^2+\left(-\frac{21}{128}  x_2+\frac{5}{768}  x_2^3\right) x_1^3 \nonumber \\
&+& \left(\frac{185 x_2}{1536}+\frac{5 x_2^3}{1024}\right)   x_1^4 + \dots
\eea
which reproduces the analogous formula in \cite{Aganagic:2002wv}, modulo the ambiguity in the degree-zero contribution.

\subsection{Local $\mathbb{F}_2$ at large radius} 
\begin{figure}[!b]
\centering
\includegraphics{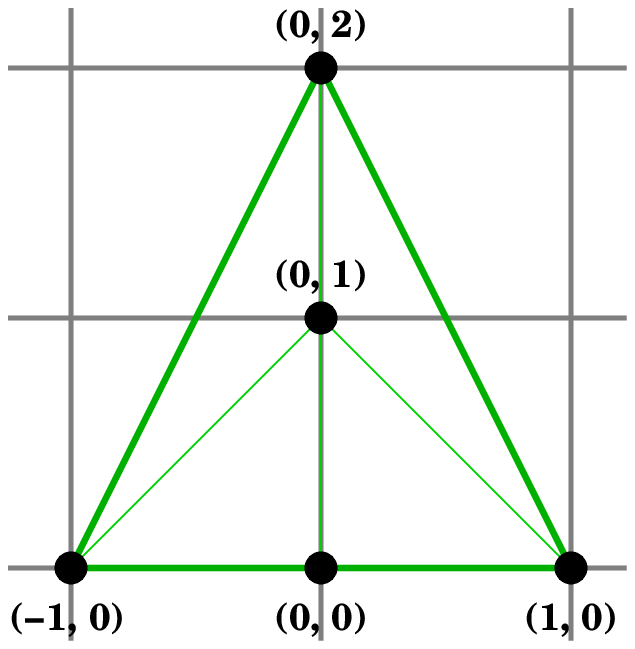}
\caption{The fan of local $\mathbb{F}_2$.}
\label{fanf2}
\end{figure}

We might proceed along the same lines for the case of local $\mathbb{F}_2$. The curve is given by

\beq
\label{curvaf2}
a_1 v + \frac{a_2}{v}= a_3 + a_4 u + a_5 u^2 
\eeq
Branch points are located at 
\beq
u + \frac{a_{4}}{2a_5} = \left\{
\begin{array}{c}
\pm \frac{\sqrt{a_{4}^2-4 a_{3} a_{5}-8 \sqrt{a_{1}} \sqrt{a_{2}} a_{5}}}{2 a_{5}} \equiv \pm c_1 \\
\pm \frac{\sqrt{a_{4}^2-4 a_{3} a_{5}+8 \sqrt{a_{1}} \sqrt{a_{2}} a_{5}}}{2 a_{5}} \equiv \pm c_2 
\end{array}\right.
\label{branchf2hom}
\eeq
and we have accordingly
\beq
\label{PiAF2}
\frac{\partial \Pi_A}{\partial a_4} = \int_{c_1}^{c_2}\frac{d X}{(X^2-c_1^2)(X^2-c_2^2)}=\frac{K\left(1-\frac{c_{2}^2}{c_{1}^2}\right)}{c_{1}}
\eeq
\beq
\label{PiBF2}
\frac{\partial \Pi_B}{\partial a_4} = \int_{-c_1}^{c_1}\frac{d X}{(X^2-c_1^2)(X^2-c_2^2)}=\frac{2 K\left
(\frac{c_{1}^2}{c_{2}^2}\right)}{c_{2}}
\eeq
In this case good coordinates associated to the base $\mathbb{P}^1$ and the $\mathbb{P}^1$ fiber are
\beq
\label{largradcoord}
z_B=\frac{a_1 a_2}{a_3^2}, \qquad z_F=\frac{a_3 a_5}{a_4^2}
\eeq
Upon setting $a_1=a_2$, $a_3=a_5=1$, periods take the form
\bea
\frac{\partial t_F}{\partial z_F} &=&\frac{\partial \Pi_A}{\partial z_F} = -\frac{2 K\left(-\frac{16 \sqrt{z_B} z_F}{-8 \sqrt{z_B} z_F-4 z_F+1}\right)}{\pi  z_F \sqrt{1-4 \left(2
   \sqrt{z_B}+1\right) z_F}} \nonumber \\
\frac{\partial^2 \mathcal{F}}{\partial z_F\partial t_F} &=& \frac{\partial \Pi_B}{\partial z_F} = -\frac{4 K\left(\frac{-8 \sqrt{z_B} z_F-4 z_F+1}{8 \sqrt{z_B} z_F-4 z_F+1}\right)}{z_F \sqrt{1-4 \left(1-2
   \sqrt{z_B}\right) z_F}} \nonumber \\
t_B &=& \Pi_{0_+} - \Pi_{0_-} =2 i \tan ^{-1}\left(\sqrt{4 z_B-1}\right)
\eea
where the normalization has been chosen in order to get the right asymptotics. Integration and inversion yields the mirror map at the large radius point
\bea
z_B(Q_B)&=&\frac{Q_B}{(Q_B+1)^2} \nonumber \\
z_F(Q_B,Q_F)&=&\left(1+Q_B\right) Q_F+\left(-2-4 Q_B-2 Q_B^2\right) Q_F^2\nonumber \\
&+& \left(3+3 Q_B+3
   Q_B^2+3 Q_B^3\right) Q_F^3 + \dots
\label{mirrormapf2}
\eea
with $Q_B=e^{-t_B}$, $Q_F=e^{-t_F}$ and therefore

\bea
\partial_{t_F}\mathcal{F}(Q_B,Q_F)&=&
\left(\log (Q_F) \log (Q_B Q_F)\right)+\left(4+4 Q_B\right) Q_F+ (1+16
   Q_B+ \nonumber \\ 
   &+& Q_B^2) Q_F^2 + \left(\frac{4}{9}+36 Q_B+36 Q_B^2+\frac{4
   Q_B^3}{9}\right) Q_F^3 \nonumber \\
&+&\left(\frac{1}{4}+260 Q_B^2+64 (Q_B+Q_B^3)\right)
   Q_F^4+
\dots
\eea
as in \cite{Chiang:1999tz}.
\section{Local $\mathbb{F}_2$ and  $[\mathbb{C}^3 / \mathbb{Z}_4]$ orbifold Gromov-Witten invariants}
\subsection{Orbifold mirror map and genus zero invariants}
\begin{figure}[!t]
\centering
\includegraphics{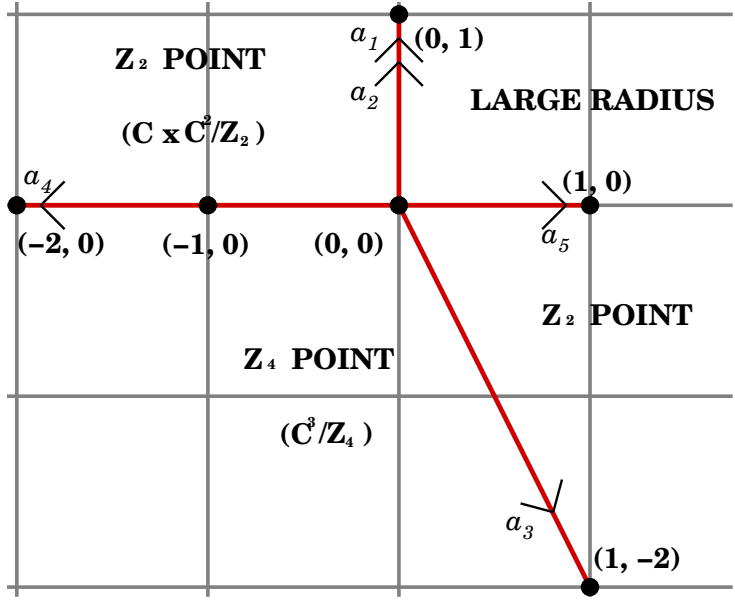}
\caption{The secondary fan of local $\mathbb{F}_2$.}
\label{secfanf2}
\end{figure}

We will now apply the considerations above to the study of the tip of the classical K\"ahler moduli space for local $\mathbb{F}_2$, where the compact divisor collapses to zero size. The resulting geometry \cite{Benvenuti:2004dy} is a $\mathbb{Z}_4$ orbifold of $\mathbb{C}^3$ by the action $(\omega; z_1,z_2,z_3) \to (\omega z_1, \omega z_2, \omega^{-2} z_3)$, with $\omega \in \mathbb{Z}_4$. In the orbifold phase, the genus zero closed amplitude computes \cite{Zaslow:1992rp} the generating function of genus-zero correlators of twist fields
\beq
\mathcal{F}^{orb}(s_{1/4}, s_{1/2})=\sum_{n,m} \frac{1}{n! m!} \langle \mathcal{O}_{1/4}^m \mathcal{O}_{1/2}^n \rangle s_{1/4}^m s_{1/2}^n 
\label{preptwist}
\eeq
In (\ref{preptwist}) the sum is over the generators $s_{1/4}$, $s_{1/2}$ of the orbifold cohomology ring and they are associated respectively with the twisted sectors $1/4$ and $1/2$ under the $\mathbb{Z}_4$ action. The corresponding topological observables are denoted respectively as $\mathcal{O}_{1/4}$ and $\mathcal{O}_{1/2}$. The correlators $\langle \mathcal{O}_{1/4}^m \mathcal{O}_{1/2}^n \rangle$ compute genus-zero {\it orbifold Gromov-Witten invariants} $N^{orb}_{0, (m,n)}$ with $m$ insertions of weight $1/4$ and $n$ of weight $1/2$.
\begin{figure}[!b]
\centering
\label{fanc3z4}
\includegraphics{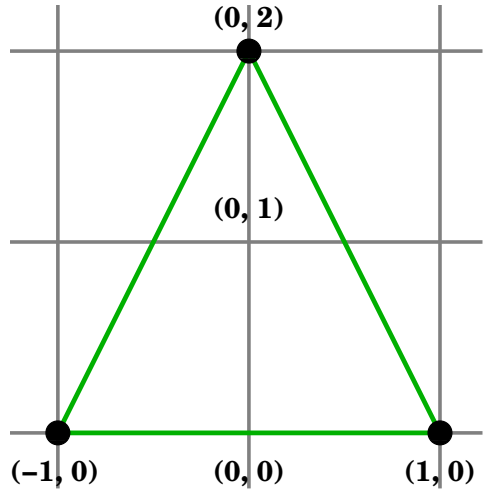}
\caption{The fan of $[\mathbb{C}^3/\mathbb{Z}_4]$.}
\end{figure}
\\From figure \ref{fanf2} we see that Mori vectors for local $\mathbb{F}_2$ are

\beq
\label{glsmf2}
\begin{array}{ccccccc}
Q_1 &=& (0,& 1,& 1,& 0,& -2) \\
Q_2 &=& (1,& -2,& 0,& 1,& 0)
\end{array}
\eeq
and the mirror curve $\Gamma_{2,2}$ has the form (\ref{curvaf2})

$$a_1 v + \frac{a_2}{v}= a_3 + a_4 u + a_5 u^2 $$
Following \cite{DelaOssa:2001xk} we argue that the point we are looking for in the $B$-model moduli space is given by $a_3=a_4=0$. This would amount to shrinking to zero size the compact divisor given, 
in the homogeneous coordinates $z_i$ introduced in section \ref{sectypq}, by $z_5=0$. When resolving $\mathbb{C}^3/\mathbb{Z}_4$, the latter corresponds to the extra divisor in the blow-up procedure: indeed, dropping $z_5$ from the $GLSM$ (\ref{glsmf2}) leads one to the system of charges of the base $\mathbb{F}_2$ inside local $\mathbb{F}_2$. \\ This argument is strengthened by the following remark. The secondary fan of (\ref{glsmf2}) is shown in figure \ref{secfanf2} and has the set of charges (see (\ref{secondaryglsm}))
\beq
\label{glsmsecfanf2}
\begin{array}{c|cccccccc}
&  a_1 & a_3 & a_{5} & a_2 & a_4 \\
\hline
Q_1 & 1,& 0,& 0,& -1,& 0 \\
Q_2 & 0,& 0, & 2, & 0, & 1 \\
Q_3 & 1, & 1, & 1, & 1, & 1
\end{array}
\eeq
The fan of the toric compactification $\overline{\mathcal{M}^{B,tor}_{2,2}}$ of $\mathcal{M}^B_{2,2}$ is simplicial but with marked points: $\overline{\mathcal{M}^{B,tor}_{2,2}}$ is thus a toric orbifold. Its orbifold patches are, as shown in figure \ref{secfanf2}, a smooth $\mathbb{C}^2$ patch containing the large complex structure point, two non-smooth $[\mathbb{C}^2/\mathbb{Z}_2]$ cones, and finally a $[\mathbb{C}^2/\mathbb{Z}_4]$ patch parameterized by $(a_3,a_4)$. Inspection shows that the latter is a toric orbifold of $\mathbb{C}^2$ by the action

\beq
\label{actionz4}
\begin{array}{ccccc}
\mathbb{Z}_4 & \times & \mathbb{C}^2 &\to & \mathbb{C}^2 \\
\lambda & & (x,y) & \to & (\lambda x, \lambda^2 y)
\end{array}
\eeq
$(a_3,a_4)=(0,0)$ is therefore the only $\mathbb{Z}_4$ point in the compactified moduli space as expected. From (\ref{actionz4}) we see that good coordinates around $(a_3,a_4)=(0,0)$ are given by 

\bea
a_3 &=& \sqrt{d} e \nonumber \\
a_4 &=& d^{1/4}
\eea
Let us then find a complete basis of solutions for the $GKZ$ system around this point. Picard-Fuchs operators are written in this patch as

\bea
\label{PFf2}
\mathcal{L}_1 &=& a_3 \partial^2_{a_4}+\frac{1}{2}\theta_{a_4}\theta_{a_3} \nonumber \\
\mathcal{L}_2 &=& \partial^2_{a_3}-\frac{1}{16}(\theta^2_{a_4}-4\theta^2_{a_3})-\frac{1}{4}\theta_{a_3}\theta_{a_4}
\eea
and the branch points (\ref{branchf2hom}) here read 

\bea
\pm c_1 &=& \pm \frac{1}{2} \sqrt{a_{4}^2-4 a_{3}-8} \nonumber \\
\pm c_2 &=& \pm \frac{1}{2} \sqrt{a_{4}^2-4 a_{3}+8}
\eea
while the period integrals (\ref{PiAF2}),(\ref{PiBF2}) and (\ref{residuolog}) become

\bea
\label{perF2}
\partial_{a_4}\Pi_A &=&
\frac{K\left(\frac{a_4^2-4 a_3-8}{a_4^2-4 a_3+8}\right)}{\sqrt{a_4^2-4 a_3+8}}-\frac{K\left(\frac{a_4^2-4
   a_3+8}{a_4^2-4 a_3-8}\right)}{\sqrt{a_4^2-4 a_3-8}} \nonumber \\
\partial_{a_4}\Pi_B &=& 
2 \frac{K\left(\frac{a_4^2-4 a_3-8}{a_4^2-4 a_3+8}\right)}{\sqrt{a_4^2-4 a_3+8}} \nonumber \\
\Pi_{0_\pm} &=& \log \left(\frac{a_3}{2}\pm\frac{\sqrt{a_3^2-4}}{2}\right)
\eea
We want to find solutions of the $PF$ system (\ref{PFf2}) with prescribed monodromy around $(d,e)=(0,0)$, in order to match them with the conjugacy classes of $\mathbb{Z}_4$. Defining 

\bea
\label{cambvars14}
s_{1/4} &=& 
\left[\frac{(8-8i) \pi^{1/2}}{\Gamma \left(\frac{1}{4}\right)^2}\right]
\left(\frac{\Pi_B}{2}-\frac{\Pi_A}{1-i}\right) \\
s_{3/4} &=& 
\left[\frac{(4+4 i) \Gamma \left(\frac{1}{4}\right)^2}{\pi ^{3/2}}\right]
\left(\frac{\Pi_B}{2}+\frac{\Pi_A}{1+i}\right) \\
\label{cambvars34}
s_{1/2}&=& -2i\Pi_{0_-}+\pi
\eea
we then have

\bea
\label{mirrormaporb14}
s_{1/4}(d,e)&=& d^{1/4}\left[1+\left(\frac{e^2}{32}-\frac{e}{192}+\frac{1}{2560}\right) d -\frac{25 e^3}{18432} d^2 + \dots\right] \\ 
\label{mirrormaporb12}
s_{1/2}(d,e)&=& d^{1/2}\Bigg[
e+\frac{e^3 d}{24}+\frac{3 e^5 d^2}{640}+\frac{5 e^7 d^3}{7168}+\dots\Bigg] 
\\
s_{3/4}(d,e)&=&
d^{3/4}\left[\left(e-\frac{1}{12}\right)+\left(\frac{3 e^3}{32}-\frac{3 e^2}{128}+\frac{9 e}{2560}-\frac{3}{14336}\right) d +\dots\right] 
\eea
The normalization of the mirror map has been fixed by imposing the correct asymptotics $s_{1/4}\sim a_4$, $s_{1/2}\sim a_3$ as to reproduce the generators of the classical orbifold cohomology. These are given by $a_4$ and $a_3$ respectively for the weight $1/4$ and $1/2$ twisted sectors. Concerning the solution $s_{3/4}$, this is identified with the derivative of the generating function $\mathcal{F}_{orb}$ in (\ref{preptwist}) with respect to $s_{1/4}$; in fact, the orbifold cohomology pairing modifies this relation by a factor of $4$, i.e. $s_{3/4}=4\partial_{s_{1/4}}\mathcal{F}_{orb}$. Taking all this into account, inversion of (\ref{mirrormaporb14}) and (\ref{mirrormaporb12}) gives the following expression for the prepotential
\bea
4\frac{\partial\mathcal{F}_{orb}}{\partial s_{1/4}}(s_{1/2}, s_{1/4}) &=&
\left(s_{1/2}+\frac{s_{1/2}^3}{48}+\frac{s_{1/2}^5}{960}+\frac{29 s_{1/2}^7}{430080}+\frac{457
   s_{1/2}^9}{92897280}+O\left(s_{1/2}^{11}\right)\right) s_{1/4} \nonumber \\
&+& \left(-\frac{1}{12}-\frac{s_{1/2}^2}{96}-\frac{11 s_{1/2}^4}{9216}-\frac{49
   s_{1/2}^6}{368640}-\frac{601 s_{1/2}^8}{41287680}+O\left(s_{1/2}^{10}\right)\right) s_{1/4}^3 \nonumber \\
&+&\left(\frac{7
   s_{1/2}}{3840}+\frac{s_{1/2}^3}{1920}+\frac{47 s_{1/2}^5}{460800}+\frac{6971 s_{1/2}^7}{412876800}+O\left(s_{1/2}^{9}\right)\right) s_{1/4}^5 \nonumber \\
&+& \dots 
\eea
As a check, the prepotential thus obtained is invariant under monodromy. The first few orbifold GW invariants
are listed in table \ref{tableorb}. Our predictions exactly match the results\footnote{We 
are grateful to Tom Coates for sharing with us his computations and for enlightening discussions on this point.} obtained in \cite{coates}  
after the methods of \cite{cclt}.
\begin{table}[t]
\centering
\begin{tabular}{|c|cccccc|}
\hline
 & $m$ & 2 & 4 & 6 & 8 & 10\\
\hline
$n$ & &  & & & & \\

0& &  0 & $-\frac{1}{8}$ & 0 & $-\frac{9}{64}$ & 0\\
1& & $\frac{1}{4}$ & 0 & $\frac{7}{128}$ & 0 & $\frac{1083}{1024}$\\
2& & 0 & $-\frac{1}{32}$ & 0 & $-\frac{143}{512}$ & 0\\
3& & $\frac{1}{32}$ & 0 & $\frac{3}{32}$ & 0 & $\frac{85383}{16384}$\\
4& & 0 & $-\frac{11}{256}$ & 0 & $-\frac{159}{128}$  & 0\\
5& & $\frac{1}{32}$ & 0 & $\frac{47}{128}$ & 0 & $\frac{360819}{8192}$\\
6& & 0 & $-\frac{147}{1024}$ & 0 & $-\frac{157221}{16384}$  & 0\\
7& & $\frac{87}{1024}$ & 0 & $\frac{20913}{8192}$ & 0 & $\frac{73893099}{131072}$\\
8& & 0 & $-\frac{1803}{2048}$ & 0 & $-\frac{3719949}{32768}$  & 0\\
9& & $\frac{457}{1024}$ & 0 & $\frac{1809189}{65536}$ & 0 & $\frac{5312434641}{524288}$\\
10& & 0 & $-\frac{70271}{8192}$ & 0 & $-\frac{498785781}{262144}$  & 0\\
11& & $\frac{7859}{2048}$ & 0 & $\frac{56072653}{131072}$ & 0 & $\frac{254697581847}{1048576}$\\
12& & 0 & $-\frac{15933327}{131072}$ & 0 & $-\frac{11229229227}{262144}$  & 0\\
13& & $\frac{801987}{16384}$ & 0 & $\frac{2354902131}{262144}$ & 0 & $\frac{31371782305803}{4194304}$\\
& & & & & & \\
\hline
\end{tabular}
\caption{Genus zero orbifold Gromov-Witten invariants $N^{orb}_{0, (m,n)}$ of $[\mathbb{C}^3/\mathbb{Z}_4]$.}
\label{tableorb}
\end{table}

\subsection{Adding D-branes}
\begin{figure}[t]
\centering
\includegraphics{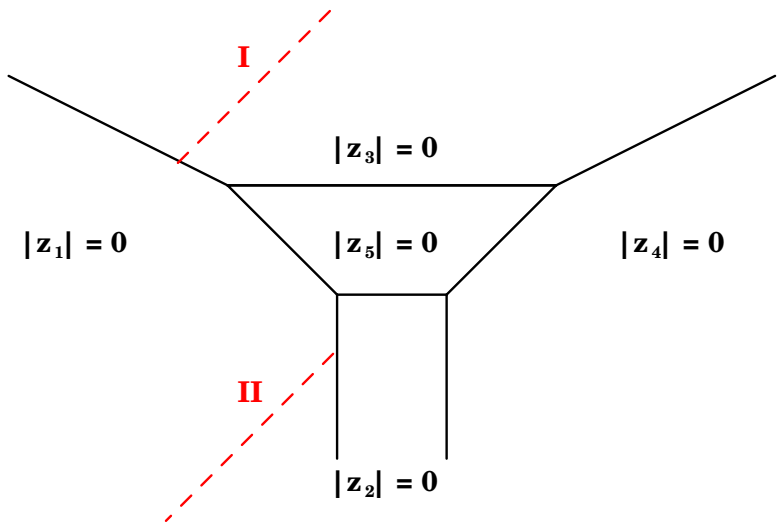}
\caption{The pq-web of local $\mathbb{F}_2$ with lagrangian branes on an upper (I) and lower (II) outer leg.}
\label{f2brane}
\end{figure}
Following the discussion of section \ref{openstrings} we might want to turn on an open sector and add 
Lagrangian branes to the orbifold. 
The procedure of \cite{Aganagic:2000gs, Aganagic:2001nx} is in principle valid away from the region of semi-classical 
geometry and has had a highly non-trivial check for the local $\mathbb{F}_0$ case
in \cite{Bouchard:2007ys}, where open amplitudes have been matched against Wilson lines in the large $N$ dual Chern-Simons theory. 
First of all, we will consider the setups I and II of figure \ref{f2brane}, with a $D$-brane ending respectively on the outer leg $|z_1|=|z_3|$ and $|z_1|=|z_2|$. The choice of variables (\ref{polynomialpq}) we have made for the mirror curve $\Gamma_{2,2}$, in which the $B$-model coordinate mirror to $|z_2|$ was gauge-fixed to one, corresponds to phase $II$. This means that $v$ is the
variable that goes to one on the brane and $X_{(II)}\equiv u$ is the good open string parameter to be taken as the 
independent variable in (\ref{openflat}) \cite{Bouchard:2007ys}. The transition from phase II to phase I is accomplished by the (exponentiated) $SL(2,\mathbb{Z})$ transformation
\bea
X_{(II)}\equiv u  &\to & \frac{1}{u} \equiv X_{(I)} \nonumber \\
v &\to & v u^2 
\eea
Accordingly, the differential (\ref{differpq}) has the form
\beq
d\lambda=\left\{ \begin{array}{cc}\log{\frac{a_3 X_{(I)}^2 + a_4 X_{(I)} + 1+\sqrt{(a_3 X_{(I)}^2 + a_4 X_{(I)} + 1)^2-4 X_{(I)}^4}}{2 X_{(I)}^4}}\frac{d X_{(I)}}{X_{(I)}} & \hbox{phase  I}\\ \log{\frac{a_3 + a_4 X_{(II)} + X_{(II)}^2+\sqrt{(a_3 + a_4 X_{(II)} + X_{(II)}^2)^2-4}}{2}}\frac{d X_{(II)}}{X_{(II)}} & \hbox{phase  II}\end{array}\right.
\eeq
We now turn to analyze the unframed $A$-model disc amplitude for a brane in phase I. In order to do that we have to compute the instanton corrected open modulus (\ref{openflat}) and the Abel-Jacobi map (\ref{abeljac}). To determine the former, and more precisely the $r_i$ coefficients in (\ref{openflat}), we use the result of \cite{Lerche:2001cw}, where the authors show that for this outer-leg configuration the large radius open flat variable solving the extended Picard-Fuchs system is given by
\beq
\label{zblog1}
z^{LR}_{open} = z+\frac{t_B}{4}+\frac{t_F}{2}+\pi i
\eeq
where $z=\log{X_{(I)}}$. In the ($a_3$, $a_4$) patch containing the orbifold point this becomes
\beq
\label{zblog2}
z^{LR}_{open} = z+\pi i + \mathcal{O}(a_4) + \mathcal{O}(a_3)
\eeq
Notice that in (\ref{zblog1}), (\ref{zblog2}), both $z^{LR}_{open}$ and $z+\pi i$ solve the extended $PF$ system and can then serve as a flat coordinate: $z^{LR}_{open}$ does the job by construction, and the same is true for $z$ because it is a difference of solutions of the Picard-Fuchs system by (\ref{zblog1}). Following \cite{Bouchard:2007ys}, we have that the difference $z^{LR}_{open} -\frac{t_B}{4}-\frac{t_F}{2}=z+\pi i$ is a global open flat variable and serves as the expansion parameter at the orbifold point. In terms of exponentiated variables, we then have:
\beq
\label{ZB}
Z^{orb}_{open}= - X_{(I)}
\eeq
Having the mirror map and using (\ref{abeljac}) or (\ref{incompletelauricella}) one can then mimic \cite{Bouchard:2007ys} and expand the chain integral, thus obtaining the disc amplitude $\mathcal{F}_{0,1}(a_3,a_4,z)$ as a function of the bare variables, or, using (\ref{mirrormaporb14})-(\ref{mirrormaporb12}), of the flat variables. Notice that, since $(a_3,a_4)$ have non-trivial $\mathbb{Z}_4$ transformations, in order to preserve the fact that the curve (\ref{curvaf2}) stays invariant we are forced to assign weights $(1/4,1/2)$ to $(u,v)$ respectively, and so according to (\ref{ZB}) $Z^{orb}_{open}$ has weight $-1/4$. Eventually we get
\begin{eqnarray}
\mathcal{F}_{0,1}(s_{1/4}, s_{1/2}, Z^{orb}_{open}) &=& 
\left(-\frac{s_{1/2}    s_{1/4}^3}{192}+\frac{s_{1/2}^2 s_{1/4}}{32}-s_{1/4}\right) Z^{orb}_{open} \nonumber \\
&+&\left(\frac{s_{1/2}^2    s_{1/4}^2}{64}-\frac{s_{1/4}^2}{4}+s_{1/2} \left(\frac{1}{2}-\frac{s_{1/4}^4}{384}\right)\right) (Z^{orb}_{open})^2 \nonumber \\
&+&\left(\frac{7 s_{1/2}^2 s_{1/4}^3}{576}-\frac{s_{1/4}^3}{9}+\frac{s_{1/2} s_{1/4}}{3}\right) (Z^{orb}_{open})^3
\nonumber \\ &+& \dots
\label{f01}
\end{eqnarray}
which is monodromy invariant. The amplitude (\ref{f01}) should correspond to a generating function
of open Gromov-Witten invariants of the $\mathbb{C}^3/\mathbb{Z}_4$ orbifold.\\
The situation for phase II appears to be more subtle.  The resulting
topological amplitude computed from the chain integral (\ref{abeljac}) picks up a sign flip under $\mathbb{Z}_4$. This is not completely surprising,
since it is known that disc amplitudes may have non-trivial monodromy \cite{ov}, and it might also be seen to be related to the more complicated geometrical structure of the $\mathbb{Z}_4$ orbifold with respect to the $\mathbb{Z}_3$ case, due to the presence of non-trivial stabilizers for the cyclic group action.

\subsection{Modular structure of topological amplitudes}

Higher genus amplitudes are associated to the quantization of the symplectic space spanned by the periods of the mirror
curve \cite{Witten:1993ed}. The corresponding topological 
wave functional obeys recursion relations (BCOV equations \cite{bcov}) that allow to compute higher genus amplitudes
building on genus zero and one results, up to holomorphic ambiguities. It has been shown in \cite{Aganagic:2006wq}
that this algorithm is made simpler and more efficient by exploiting modular properties of the topological amplitudes.

Let us summarize very briefly the results of \cite{Aganagic:2006wq} relevant for our discussion. 
As recalled in Sect. 3, the choice of $B$-model complex structures can be parametrized in terms of the periods
of the three-form $\Omega$ in a chosen symplectic basis $A^i \cap B_j = \delta^i_j$ 
in $H_3(\hat X,\mathbb{Z})$, which define a so-called ``real polarization''. 
Special geometry
relations between the periods $x^i = \int_{A^i} \Omega$ and $p_j = \int_{B_j}\Omega$ are summarized in terms 
of a prepotential ${\cal F}_0 (x^i)$ which turns out to be the genus zero free-energy of the topological string.
The ``phase space'' $(x^i,p_j)$ can be endowed with a natural symplectic structure with symplectic
form $dx^i\wedge dp_i$. The higher genus amplitudes ${\cal F}_g$ are associated to the quantization
of this space, with the string coupling $g_s^2$ playing the r\^ole of $\hbar$. More precisely, the
full topological string partition function $Z(x^i)\sim \exp{\sum_g g_s^{2g-2} F_g (x^i)}$
is interpreted as a wave function \cite{Witten:1993ed}.
The periods $(x^i,p_j)$ generically undergo an $Sp(2 p-2, \mathbb{Z})$ transformation under a change of symplectic
basis of the mirror curve. Correspondingly, the $B$-model topological amplitudes ${\cal F}_g$ have definite
transformation properties that can be derived by implementing the canonical transformation at the quantum
level on the topological wave function.
The crucial observation of \cite{Aganagic:2006wq} is that there is a finite index subgroup 
$\Gamma \subset Sp(2 p-2, \mathbb{Z})$ which is a \textit{symmetry} of the theory.
$\Gamma$ is precisely the group generated by the monodromies of the periods, which must leave invariant the
topological wave-function. This symmetry constrains the topological amplitudes; in particular in the real
polarization the ${\cal F}_g$ can be shown to be quasi-modular forms of $\Gamma$ \cite{Aganagic:2006wq},
namely they transform
with a shift. For example for the case of 
elliptic mirror curves, i.e. local surfaces, this amounts to say that the wave-function is a finite power 
series in the second Eisenstein series.

We recall that one could also have chosen to parameterize the $B$-model moduli space with
the Hodge decomposition of $H^3(\hat X,\mathbb{Z})$ in terms of a fixed background complex structure.
The topological wave function in this holomorphic polarization can be shown \cite{Witten:1993ed} to obey the BCOV
holomorphic anomaly equations. The topological amplitudes $\hat{\cal F}_g$ in this case
turn out to be proper modular forms of weight zero
under $\Gamma$, namely invariant under $\Gamma$, but they are non-holomorphic.
For elliptic mirror curves, they can be written in terms of a polynomial in a canonical, non-holomorphic extension of 
the second Eisenstein series
\beq
E_2(\tau) \to \hat{E_2}(\tau, \bar \tau) := E_2(\tau) -\frac{3}{\pi}\frac{1}{\Im m \tau}
\label{e2nonhol}
\eeq
with coefficients in the ring of holomorphic modular forms of $\Gamma$. Thus one can pass from the real
to the holomorphic polarization just by the above shift of variables. 

The advantage of the approach proposed in \cite{Aganagic:2006wq} is twofold. On one side
it simplifies the solution of BCOV equations by restricting the functional dependence
of the $\hat{\cal F}_g$ to the ring of $\Gamma$ modular forms. On the other it allows to
relate the topological amplitudes in different patches of the $B$-model moduli space
allowing in this way to extract enumerative invariants {\it e.g.} at the orbifold point.

As we will show in the following, our method is perfectly tailored to display the modular symmetry of the topological wave-function.
In fact, the relation with the Seiberg-Witten curves greatly simplifies the analysis of the modular properties
of higher genus amplitudes. Moreover, since we obtain explicit expressions for the periods of the mirror curve
in terms of the branch points, it is enough to write the latter in terms of modular forms to make manifest the
modular properties of genus zero and
one topological amplitudes, thus providing the building blocks for the solution of BCOV equations. About the latter we point out
however that there is a caveat: for the geometries under our study
in addition to the modular dependence there is also a dependence on an extra parameter (independent of $\tau$), as in the discussion of \cite{Aganagic:2006wq} about the similar case of local $\mathbb{F}_0$.
 This makes the solution of the BCOV equations at higher genus more involved computationally, since one has to fix a functional dependence
on an extra datum. We choose to handle this problem with the approach developed in
\cite{Eynard:2004mh,Bouchard:2007ys} in which the holomorphic ${\cal F}_g$ are defined 
via recurrence relations inspired by matrix-model techniques. This will allow us to display the general modular structure of the free energies in the local $\mathbb{F}_2$ case, in a way in which both the dependence on the modular variable and that on the extra parameter are completely fixed.

In this section we first find the relevant change of basis from
large radius to the orbifold point and then identify the ring of modular functions relevant for the local
$\mathbb{F}_2$ case. These results provide the necessary tools to discuss higher genus invariants, which
will be the subject of the next section.
 
\subsubsection{The change of basis from large radius}
We already saw in the last section that the mirror map at the orbifold point is obtained by choosing solutions of the $GKZ$ system which diagonalize the monodromy of the periods. This implies that the solutions at large radius ($1, t_B, t_F, \partial_F \mathcal{F}$) are related to those at the orbifold point ($1, s_{1/4},s_{1/2},s_{3/4}$) by a linear transformation, which, for the subsector relating
($t_F$, $\partial_F\mathcal{F}$) and ($s_{1/2},s_{3/4}$)  might be regarded as an (unnormalized)\footnote{The determinant of the change of basis is equal to 2, see (\ref{matrS}), due to the fact that we are dealing
with a local threefold.} automorphism in $H_1(\Gamma_{2,2},\mathbb{Z})$. 
In \cite{Aganagic:2006wq} it was showed that under a symplectic change of basis
\beq
\label{symplchange}
\vec{\Pi} \to S \vec{\Pi} = \left( 
\begin{array}{cc}
A & B \\
C & D
\end{array}
\right) \vec{\Pi}, \qquad S\in Sp(2p-2,\mathbb{Z})
\eeq
the genus-$g$ amplitudes $\mathcal{F}_g$
are subject to a transformation which can be derived 
by implementing the canonical transformation associated to (\ref{symplchange}) in the path integral 
defining the topological wave function. From saddle point expansion one then gets
\beq
\label{trasffg}
{\tilde {\cal F}}_g ={\cal F}_g + \Gamma_g(\Delta,{\cal F}_{r<g}))
\eeq
where $\Gamma_g$ is determined by the Feynman rules in terms of lower genus vertices $\partial^n {\cal F}_{r<g}$
and the propagator $\Delta$ given by 
\beq
\Delta \simeq \frac{1}{\tau + C^{-1} D}
\label{feyn}
\eeq
up to normalization factors. In (\ref{feyn}) $\tau$ is the period matrix of $\overline{\Gamma}_{2,2}$.   
This means, for instance, that knowledge of $\mathcal{F}_g$ in the large radius region allows 
one to compute genus-$g$ free energies in the full moduli space, 
provided that we know how the periods are transformed when going from one region to another. 
This was successfully exploited in \cite{Aganagic:2006wq} to predict higher genus orbifold 
Gromov-Witten invariants for the $[\mathbb{C}^3/\mathbb{Z}_3]$ orbifold from the ones of the
large radius $K_{\mathbb{P}^2}$ geometry. \\
We underline that our method of solving the extended GKZ system in the full
moduli space has the advantage to make much easier the study of the analytic continuation properties and
the consequent computation of the linear change of basis (\ref{symplchange}).
Indeed, instead of performing standard (but cumbersome) multiple Mellin-Barnes transforms, 
we can easily read off what $S$ and $\Delta$ are in our case from formulae (\ref{PiAEuler}-\ref{PiBEuler}) and (\ref{perF2}). Let us define
\beq
\vec{\Pi}_{LR}=\left(
\begin{array}{c}
1 \\ t_B \\ t_F \\ \partial_F \mathcal{F}
\end{array}
\right)
\qquad 
\vec{\Pi}_{orb}=\left(
\begin{array}{c}
1 \\ s_{1/2} \\ s_{1/4} \\ s_{3/4}
\end{array}
\right)
\eeq
We can now simply compute the change of basis between the large radius and the
orbifold point by
using the Euler integral representation\footnote{Actually, the computation
is simple only when one works with the $a_4$ derivatives of the
periods and then integrates back. This gives rise to the unknown coefficients $(\alpha,\beta,\gamma,\delta)$.
A more careful inspection of the direct asymptotic expansion of the integrals (\ref{PiAEuler}), (\ref{PiBEuler}) 
would allow one to compute explicitly $\alpha$, $\beta$, $\gamma$ and $\delta$ in (\ref{matrStilde});
anyway, all this will not overly bother us, as only the $S$ subsector in (\ref{matrStilde}) will be relevant for actual computations.}   (\ref{PiAEuler}) (see also (\ref{cambvars14}-\ref{cambvars34})). We can relate $\Pi_{orb} = \tilde{S} \Pi_{LR}$ through

\beq
\label{matrStilde}
\tilde{S}=
\left(\begin{array}{cc|cc}
1 & 0 & 0 & 0 \\
\pi & -i & 0 & 0 \\
\hline
\alpha & \beta & \multicolumn{2}{c}{\multirow{2}{*}{{\it S}}} \\
\gamma & \delta & \\
\end{array}\right)
\eeq
where

\beq
S=
\left(
\begin{array}{cc}
\frac{2 \pi ^{3/2}}{\Gamma \left(\frac{1}{4}\right)^2}
&
\frac{(1-i) \sqrt{\pi }}{\Gamma \left(\frac{1}{4}\right)^2} \\
-\frac{\Gamma \left(\frac{1}{4}\right)^2}{\sqrt{\pi }}
&
\frac{\left(\frac{1}{2}+\frac{i}{2}\right) \Gamma \left(\frac{1}{4}\right)^2}{\pi ^{3/2}}
\end{array}
\right)
\label{matrS}
\eeq

\subsubsection{Local $\mathbb{F}_2$ and $\Gamma(2)$ modular forms}
\label{secmodular}
The last formula in the previous section relates the physical periods of the large radius patch to those of the orbifold patch and represents one of the main ingredients to make predictions about orbifold Gromov-Witten invariants at higher genus. 
This should be already clear at this stage from (\ref{trasffg}), but it can in fact be brought to full power due to the beautiful results of \cite{Aganagic:2006wq}.
Let us see how this works in detail for the $p=2$, $q=2$ case we have been considering, and in particular let us  figure out what the relevant modular group $\Gamma$ is in this case. In fact, we have already answered this question: as we stressed in Section \ref{legametoda}, the family of elliptic curves in this case is the \textit{same} of its field theory limit, the only thing that changes being the symplectic structure defined on the elliptic fibration, i.e., the $SW$ differential. This was already noticed in the strictly related case of local $\mathbb{F}_0$ in \cite{Aganagic:2006wq}. 
In particular we might argue as follows for the present case. By formulae (\ref{curvamirrorpq2}), (\ref{branchf2hom}) the $\Gamma_{2,2}$ family can be written as
\beq
\label{g04curve}
Y^2 = (\hat X^2-c_1^2)(\hat X^2-c_2^2)
\eeq
where we have shifted the $X$ variable in (\ref{curvamirrorpq2}) by $\hat X=X+a_4/2$. Through the following $SL(2,\mathbb{C})$ automorphism of the $\hat X$-plane
$$  \hat X = \frac{a \tilde  X + b}{c  \tilde X + d} \qquad \tilde Y = (c \tilde X + d)^2 Y $$
\beq
a=\frac{\sqrt{c_1 c_2-c_2^2}}{\sqrt{4 c_1+4 c_2}}, \quad b=\frac{c_2 (3 c_1+c_2)}{2 \sqrt{c_2 \left(c_1^2-c_2^2\right)}}, \quad c=-\frac{\sqrt{\frac{c_1^2}{c_2}-c_2}}{2 (c_1+c_2)}, \quad d=\frac{c_1+3 c_2}{2 \sqrt{c_2 \left(c_1^2-c_2^2\right)}}
\eeq
we bring (\ref{g04curve}) to the celebrated Seiberg-Witten $\Gamma(2)$-symmetric form
\beq
\tilde Y^2 = (\tilde X^2-1)(\tilde X-u)
\label{g2curve}
\eeq
where
\beq
u=\frac{c_1^2+6 c_2 c_1+c_2^2}{(c_1-c_2)^2}
\label{uSW}
\eeq
With (\ref{uSW}) at hand we can re-express the quantities computed in the previous section as $\Gamma(2)$ modular forms, whose ring is generated by the Jacobi theta functions $\theta_2(\tau)$, $\theta_3(\tau)$, $\theta_4(\tau)$, all having modular weight $1/2$. This goes as follows: the Klein invariant $j(\tau)$ of the curve (\ref{g2curve}) is rationally related to $u$ as
\beq
j(u)=64\frac{(3+u^2)^3}{(u^2-1)^2}
\label{kleininv}
\eeq
while inversion of (\ref{uSW}) gives, writing everything for definiteness in the $(a_3, a_4)$ patch,
\beq
a_4^2-a_3 = \frac{u+3}{\sqrt{2} \sqrt{u+1}}
\label{a4u}
\eeq
Combining the two formulae above and using the definition (\ref{branchf2hom}) of $c_1$, $c_2$ we can write the latter as $\Gamma(2)$ modular forms as
\beq
c_1(\tau) = 2 \frac{\theta_4^2(\tau)}{\theta_2^2(\tau)} \qquad c_2(\tau) = 2 \frac{\theta_3^2(\tau)}{\theta_2^2(\tau)}
\label{branchf2modular}
\eeq
which, being coordinates on the moduli space, are correctly modular invariant. \\ 

Given (\ref{branchf2modular}) it is then straightforward to write the building blocks of the 
BCOV recursion in terms of modular forms. According to \cite{bcov}, the recursion relies on knowledge of the Yukawa coupling $C(\tau)$ and the genus one closed amplitude $\hat{\mathcal{F}}_1(\tau, \bar \tau)$; having exact expressions for the genus zero data as functions of the branch points, one can use (\ref{branchf2modular}) to write down explicitly all the relevant quantities as modular functions. \\ Let us analyze the large radius Yukawa coupling first. We have
$$C\equiv \frac{\partial^3 \mathcal{F}}{\partial t_F^3} = \frac{4}{\pi}\left(\frac{\partial a_4}{\partial \tau}\frac{\partial t_F}{\partial a_4}\right)^{-1}$$
Using (\ref{PiAF2}) we have
\beq
\frac{\partial t_F}{\partial a_4} = \frac{K\left(1-\frac{c_{2}^2}{c_{1}^2}\right)}{c_{1}} = \frac{\pi}{4} \theta_2^2(\tau)
\label{Kmod}
\eeq
while combining (\ref{branchf2modular}) and (\ref{branchf2hom}) yields
\beq
\frac{\partial a_4}{\partial \tau} = -\frac{2^6}{a_4(\tau)} \frac{\eta^{12}(\tau)}{\theta_2(\tau)^8}
\eeq
where $\eta(\tau)$ is Dedekind's function and we have used
\beq
2 \eta^3(\tau) = \theta_2(\tau) \theta_3(\tau) \theta_4(\tau)
\eeq
besides the modular expression of $a_4$ from (\ref{branchf2hom}) 
\beq
a_4(\tau)=2 \sqrt{4 \frac{\theta_4^4(\tau)}{\theta_2^4(\tau)}+a_3+2}
\label{a4mod}
\eeq
Putting it all together we arrive at
\beq
C(\tau)= -\frac{a_4(\tau)}{64} \frac{\theta_2^6(\tau)}{\eta^{12}(\tau)}
\label{yukmod}
\eeq
Let us now address the issue of the genus 1 free energy $\hat{\mathcal{F}}_1(\tau, \bar \tau)$. For $g=1$ the holomorphic anomaly equation of \cite{bcov} reads in the local case at hand\footnote{In the following we will suppress for notational simplicity the dependence on the $t_B$ parameter, which is an auxiliary parameter entering in the definition of the differential, and write $t\equiv t_F$ for the flat coordinate coming from an actual $A-$period integration.}
\beq
\label{holanomg=1}
\partial^2_{\bar t t} \hat{\mathcal{F}}_1 (t,\bar t) = \frac{1}{2} C_{\bar t}^{tt} C_{ttt}
\eeq
where indices in (\ref{holanomg=1}) are raised with the Weyl-Petersson metric $G_{\bar t t}=2 \Im m \tau$. Rewriting everything as a function of $\tau$, $\bar \tau$ (\ref{holanomg=1}) integrates immediately to
\beq
\label{F1nonhol}
\hat{\mathcal{F}}_1(\tau,\bar \tau)=-\frac{1}{2}\log{\Im m \tau} - \log{|\psi(\tau)|}
\eeq
where we have denoted by $\psi(\tau)$ the holomorphic ambiguity at genus one. The pole structure of the amplitude fixes it uniquely; we will do it in the next section by computing explicitly its holomorphic limit.

\subsection{Higher genus amplitudes}
In this section we will examine the higher genus amplitudes for $K_{\mathbb{F}_2}$. After discussing the genus one free energy, we will turn to the analysis of the $g>1$ closed amplitudes, treating in detail the case $g=2$, and we will give predictions for orbifold Gromov-Witten invariants of $\mathbb{C}^3/\mathbb{Z}_4$ for $g=1,2$. 
\subsubsection{One loop partition function for $Y^{p,q}$ and genus $1$ orbifold GW}
Let us address the issue of the genus 1 free energy in slightly larger generality for the full $Y^{p,q}$ class. In this latter case it would be natural to guess that, again, the same structure as in $SU(p)$ Seiberg-Witten theory holds: monodromy invariance requires it to be written in terms of $\Gamma\subset Sp(2p-2,\mathbb{Z})$ modular forms, and this would lead to the appearance of Siegel modular forms with characteristic. However we happen to have already largely answered this question in the language of hypergeometric functions of the branch points $b_i$. Indeed, denoting collectively with $\vec t$ the set of flat coordinates, the holomorphic $\vec{\bar t}$ limit of $\hat{\mathcal{F}}(\vec t, \vec{ \bar t})$ is given on general grounds \cite{bcov} as

\beq
\label{F1branch}
\mathcal{F}_1(\vec t) = -\frac{1}{2} \log\mathcal{\det{J}}-\frac{1}{12} \log{\Delta}
\eeq
where $\mathcal{J}$ is the Jacobian matrix of the $A$-periods (in the appropriate polarization) with respect to the bare variables and $\Delta$ is a rational function of the branch points, with zeroes at the discriminant locus of the curve. But it turns out that the awkward  Jacobian $\mathcal{J}$ in (\ref{F1branch})
 is precisely the main object for which we have found a closed form expression in (\ref{derperiodpq})! This is definitely an advantage with respect to finding a hyperelliptic generalization of the modular symmetry of local surfaces, and in the elliptic case it compendiates nicely the modular expression obtained in the previous section. 
In the following we will therefore use (\ref{F1branch}) in the local $\mathbb{F}_2$ case to obtain a closed expression for $\mathcal{F}_1$ in homogeneous (bare) coordinates, and then exploit (\ref{trasffg}) to compute genus 1 orbifold GW invariants of $\mathbb{C}^3/\mathbb{Z}_4$. 
\\ 
\\
From the considerations above we have at large radius
\beq
\label{F1branchLR}
\mathcal{F}_1^{LR}(t_F, t_B) = -\frac{1}{2} \log\left(\frac{\partial t_F}{\partial a_4}\right)+ \log{\left[c_1^a c_2^b (c_1 - c_2)^c (c_1 + c_2)^d\right]}
\eeq
where the exponents of the second term are fixed by the topological vertex computation as $a=-1/6$, $b=-1/6$, $c=-1/12$, $d=-1/12$.
Then, from (\ref{branchf2hom})  and (\ref{PiAF2})
\beq
\mathcal{F}_1^{LR}(Q_F, Q_B) = -\frac{1}{2} \frac{K\left(1-\frac{c_{2}^2}{c_{1}^2}\right)}{c_{1}}-\frac{1}{6} \log{c_1 c_2}-\frac{1}{12}\log{\left(c_1^2 - c_2^2\right)}
\label{F1hom}
\eeq
 and plugging in the mirror map (\ref{mirrormapf2}) we can straightforwardly compute
\bea
\label{F1LRexp}
\mathcal{F}_1^{LR}(Q_F, Q_B) &=& \left(-\frac{\log (Q_B)}{24} - \frac{\log (Q_F)}{12} \right) - \frac{Q_F}{6} - \frac{Q_F^2}{12} - \frac{Q_F^3}{18} - \frac{Q_F^4}{24} +
\\ &+&
\left(-\frac{Q_F}{6} - \frac{Q_F^2}{3} - \frac{Q_F^3}{2}\right)
   Q_B + \left(-\frac{Q_F^2}{        12} - \frac{Q_F^3}{2} + \frac{37Q_F^4}{6} \right)Q_B^2 + \dots \nonumber
\eea
which is the correct form predicted by the topological vertex computation \cite{topv}. In order to verify explicitly the assertions of the previous section, we can also use the modular expression of $c_1$, $c_2$ and $\partial_{a_4} t_F$ to obtain the holomorphic limit of (\ref{F1nonhol}) in quasi-modular form. We indeed get
\beq
\mathcal{F}_1^{LR}(\tau)=-\frac{1}{2}\log{\eta(\tau)}
\label{f1eta}
\eeq
Plugging in the expression for the modular parameter $q=e^{2\pi i \tau}$ in exponentiated flat coordinates which can be computed from (\ref{mirrormapf2}),  (\ref{kleininv}) and (\ref{a4u})
\beq
q(Q_B,Q_F) =  Q_B Q_F^2+\left(4 Q_B^2+4 Q_B\right) Q_F^3+\left(10 Q_B^3+48 Q_B^2+10 Q_B\right)
   Q_F^4+O\left(Q_F^5\right)
\label{qmod}
\eeq
we recover precisely (\ref{F1LRexp}).
 \\ \\
\begin{table}[t]
\centering
\begin{tabular}{|c|ccccccc|}
\hline
 & $m$ & 0 & 2 & 4 & 6 & 8 & 10\\
\hline
$n$ & &  & & & & &\\
0 & &   & 0 & $\frac{1}{128}$ & 0 & $\frac{441}{4096}$ & 0 \\
1 & &  0 & -$\frac{1}{192}$ & 0 & -$\frac{31}{1024}$ & 0 & -$\frac{71291}{32768}$ \\
2 & & $\frac{1}{96}$ & 0 & $\frac{35}{3072}$ & 0 & $\frac{235}{512}$ & 0 \\
3 & & 0 & -$\frac{5}{768}$ & 0 & -$\frac{485}{4096}$ & 0 & -$\frac{2335165}{131072}$ \\
4 & & $\frac{7}{768}$ & 0 & $\frac{485}{12288}$ & 0 & $\frac{458295}{131072}$ & 0 \\
5 & & 0 & -$\frac{39}{2048}$ & 0 & -$\frac{40603}{49152}$ & 0 & -$\frac{58775443}{262144}$ \\
6 & & $\frac{31}{1536}$ & 0 & $\frac{2025}{8192}$ & 0 & $\frac{10768885}{262144}$ & 0 \\
7 & & 0 & -$\frac{2555}{24576}$ & 0 & -$\frac{293685}{32768}$ & 0 & -$\frac{522517275}{131072}$ \\
8 & & $\frac{2219}{24576}$ & 0 & $\frac{240085}{98304}$ & 0 & $\frac{1437926315}{2097152}$ & 0 \\
9 & & 0 & -$\frac{22523}{24576}$ & 0 & -$\frac{73017327}{524288}$ & 0 & -$\frac{397762755193}{4194304}$ \\
10 & & $\frac{16741}{24576}$ & 0 & $\frac{54986255}{1572864}$ & 0 & $\frac{32280203275}{2097152}$ & 0 \\
11 & & 0 & -$\frac{389975}{32768}$ & 0 & -$\frac{18440181205}{6291456}$ & 0 & -$\frac{12177409993695}{4194304}$ \\
12 & &  $\frac{1530037}{196608}$ & 0 & $\frac{1434341595}{2097152}$ & 0 & $\frac{7495469356455}{16777216}$ & 0 \\
& & & & & & &
\\
\hline
\end{tabular}
\caption{Genus one orbifold Gromov-Witten invariants $N^{orb}_{1,(m,n)}$ of $[ \mathbb{C}^3/\mathbb{Z}_4 ]$}
\label{genus1OGW}
\end{table}
Knowing $\mathcal{F}_1$ for local $\mathbb{F}_2$ we can straightforwardly obtain a prediction for genus one orbifold Gromov-Witten invariants of $\mathbb{C}^3/\mathbb{Z}_4$; to relate (\ref{F1LRexp}) to the expansion of the $g=1$ topological partition function at the orbifold point we just need to specialize the Feynman expansion (\ref{trasffg}) to the case at hand. The one loop term is given as
\beq
\Gamma_1=\frac{1}{2}\log{\frac{4}{\mathcal{\tau}+C^{-1} D}}
\label{Gamma1}
\eeq
where the factor of $4$ comes again from the orbifold cohomology pairing. The genus one orbifold free energy will be then given as
\beq
\mathcal{F}^{orb}_{1}=\mathcal{F}^{LR}_{1} + \Gamma_1
\label{Forb1Gamma}
\eeq
where $\tau$ can be written as a function of the $a_i$ variables using (\ref{PiAF2}), (\ref{PiBF2}) or directly as an expansion in orbifold flat coordinates using (\ref{matrS}). We can now rewrite everything in terms of $s_{1/2}$,  $s_{1/4}$ by plugging the orbifold mirror map (\ref{mirrormaporb12}) into the expression (\ref{F1branchLR}) of $\mathcal{F}_1^{LR}$, yielding
\bea
\mathcal{F}_{1}^{orb}(s_{1/4},s_{1/2}) &=&  \sum_{n,m} \frac{N^{orb}_{1,(m,n)}}{n! m!}  s_{1/4}^m s_{1/2}^n\\
&=& -\frac{s_{14}^2 s_{12}}{384}+\frac{s_{12}^2}{192}-\frac{5 s_{14}^2 s_{12}^3}{9216}+\frac{7
   s_{12}^4}{18432}-\frac{13 s_{14}^2 s_{12}^5}{163840}+\frac{31 s_{12}^6}{1105920}+\dots \nonumber
\eea
The same result would be obtained by using a modified expression for the Jacobian in (\ref{F1branch}) adapted to the orbifold patch, i.e. by replacing $\partial_{a_4} t_F$ in (\ref{F1branchLR}) with $\partial_{a_4} s_{1/4}$, without then considering the extra piece (\ref{Gamma1}) from the Feynman expansion. The first few GW invariants are reported in table \ref{genus1OGW}.

\subsubsection{Generalities on $g>1$ free energies}
Turning now to the case of $g>1$ closed amplitudes, let us first of all recall the main statements put forward 
in \cite{Aganagic:2006wq, Grimm:2007tm} for the computation of higher genus free energies. 
As mentioned in section \ref{secmodular}, modular symmetry is an amazingly stringent constraint. For 1-parameter models with elliptic mirror curve (like $SU(2)$ Seiberg-Witten theory, or local $\mathbb{P}_2$) the authors of \cite{Aganagic:2006wq, Grimm:2007tm} claim that solutions of the genus $g$ holomorphic anomaly equations can be written for $g\geq 2$ as
\beq
\hat{\mathcal{F}}_g(\tau,\bar \tau) = C^{2g-2}(\tau) \sum_{k=1}^n \hat{E}^k_2(\tau,\bar \tau)c^{(g)}_k(\tau) + C^{2g-2}(\tau) c^{(g)}_0(\tau)
\label{fgeisen}
\eeq
where $\tau$ is the complex modulus of the mirror torus, $C$ is the Yukawa coupling $\partial^3_t \mathcal{F}_{0}$, $c_k^{(g)}(\tau)$ are $\Gamma$-modular forms of weight $6(g-1)-2k$ and the full non-holomorphic dependence of $\hat{\mathcal{F}}_g$ is captured by the modular, non-holomorphic extension of the second Eisenstein series (\ref{e2nonhol}).\\


Now, there are two ways to compute the expressions in (\ref{fgeisen}). The first one consists in a direct study of the BCOV equations: in this context the holomorphic modular coefficients $c_k^{(g)}$ for $k>0$ can either be fixed  by the Feynman expansion (\ref{trasffg}) in terms of derivatives of lower genus $\mathcal{F}_{g'}$, or much more efficiently by exploiting the modular symmetry to perform a direct integration of the holomorphic anomaly equations as in \cite{Grimm:2007tm}. \\ Within this method, the only real issue is to fix the so-called ``holomorphic ambiguity'' at $k=0$, i.e. $c^{(g)}_0(\tau)$. In the 1-parameter cases analyzed in \cite{Aganagic:2006wq, Grimm:2007tm}, this is systematically done by plugging into (\ref{fgeisen}) an ansatz for $c^{(g)}_0(\tau)$ which is then determined from extra boundary data. More in detail, this works as follows: at fixed genus $g$, $c^{(g)}_0(\tau)$ is a weight $w=6g-3$ modular form\footnote{Recall that $\mathcal{F}_g(\tau,\bar \tau)$ is modular invariant and that $C(\tau)$ has weight $-3$ - see for example (\ref{yukmod}) for the local $\mathbb{F}_2$ case.}. Now, the ring of weight $w$ holomorphic modular forms $\mathcal{M}_w(\Gamma)$ is finitely generated, and the analytic behavior of $\mathcal{F}_g(\tau,\bar \tau)$ at large radius allows to write an ansatz for $c^{(g)}_0(\tau)$ with only a finite number of unknown coefficients. At the same time, $c^{(g)}_0(\tau)$ is constrained to satisfy the so-called ``gap condition'' \cite{Huang:2006si}: this imposes a sufficient number of constraints to completely determine (indeed, overdetermine) the conjectured form of the ambiguity as a function of the generators of $\mathcal{M}_w(\Gamma)$. \\ The discussion of section \ref{secmodular} has shown that the case of local $\mathbb{F}_2$ is in many ways similar to the simpler examples of $SU(2)$ Seiberg-Witten theory and local $\mathbb{P}^2$. However there is an extra complication, given by the fact that the elliptic modulus $\tau$ is not the only variable in the game: here we actually have an extra bare parameter $a_3$, or $z_B$, which is independent on $\tau$ and is related to the K\"ahler volume of the base $\mathbb{P}^1$ (see (\ref{residuolog}), (\ref{perF2})). That is, we deal here with a {\it two}-parameter model, even though with an elliptic mirror curve, and we have to properly take this into account. A first consequence of this fact is that the idea of using the gap condition to fix the holomorphic ambiguity becomes computationally more complicated, since our task is no longer reduced to fix simply a finite set of unknown \textit{numerical coefficients} of $c^{(g)}_0(\tau)$ as generated by a basis of $\mathcal{M}_w(\Gamma)$: rather we should fix a finite set of {\it unknown  functions of $a_3$}. \\

A second possibility is to avail ourselves of the framework proposed in \cite{Bouchard:2007ys} for the computation of
topological string amplitudes based on the Eynard-Orantin recursion for matrix models. This is based on a sequence of polydifferentials $W_h^{(g)}$ on the mirror curve $\Gamma$, which are recursively computed in terms of residue calculus on $\Gamma$ and out of which it is possible to extract the free energies $\mathcal{F}_g$ at any given genus.  Let us briefly review here this formalism in order to describe the general structure of higher amplitudes, referring the reader to \cite{Bouchard:2007ys, Eynard:2004mh} for further details. \\
The ingredients needed are the same as for genus zero amplitudes, namely the family of Hori-Vafa mirror curves $H(u,v)=0$  (\ref{polynomialpq}) with differential $d\lambda$ (\ref{differpq}). The genus $g$ free energies are then recursively given as
\begin{equation}
\mathcal{F}_g = {1 \over 2 - 2 g } \sum_{b_i} \underset{u=b_i}{\rm Res~} \phi(u) W_1^{(g)}(u) .
\label{rec2}
\end{equation}
where $\phi(u)$ is any antiderivative of the Hori-Vafa differential 
\beq
d \phi(u) = d \lambda(u)=\log{v(u)}\frac{du}{u}
\label{philambda}
\eeq
and $W_h^{(g)} (p_1, \ldots, p_h)$ with $g,h \in \mathbb{Z}^+$, $h \geq 1$ are an infinite sequence of meromorphic differentials on the curve defined by the Eynard-Orantin recursion
\beq
W_1^{(0)}(p_1) = 0, \qquad W_2^{(0)}(p_1,p_2) = B(p_1,p_2),
\label{w20}
\eeq

\bea
W_{h+1}^{(g)}(p, p_1 \ldots, p_h) &=& \sum_{b_i}  \underset{q=b_i}{\rm Res~} {d E_{q}(p) \over d\lambda(q) - d\lambda(\bar q)} \Big ( W^{(g-1)}_{h+2} (q, \bar q, p_1, \ldots, p_{h} )\nonumber \\
&+&\sum_{l=0}^g \sum_{J\subset H} W^{(g-l)}_{|J|+1}(q, p_J) W^{(l)}_{|H|-|J| +1} (\bar q, p_{H\backslash J}) \Big).
\label{rec1}
\eea
In the formulae above, $\bar q$ denotes the conjugate point to $q$, $B(p,q)$ is the Bergmann kernel, the one form $dE_q(p)$ is given as
\beq
dE_q(p)=\frac{1}{2}\int_q^{\bar q}B(p,\xi) d\xi
\eeq
and finally, given any subset  $J=\{i_1, \cdots, i_j\}$ of $H:=\{1, \cdots, h\}$, we defined $p_J=\{p_{i_1}, \cdots, p_{i_j}\}$. We refer the reader to \cite{Bouchard:2007ys, Eynard:2004mh} for an exhaustive description of the objects introduced above. \\

 At a computational level, the formalism of \cite{Bouchard:2007ys} is somewhat lengthier than the one of \cite{Grimm:2007tm} for computing higher genus $\mathcal{F}_g$.
On the other hand, the recursion of \cite{Bouchard:2007ys} has the great advantage of providing unambiguous 
results, 
with the holomorphic ambiguity $c^{(g)}_0(\tau)$ automatically fixed. This precisely overcomes the problem raised above. In the next section, we will therefore follow this second path to complete the discussion of section \ref{secmodular} by displaying explicitly the modular structure of the $\mathcal{F}_g$ obtained through (\ref{rec2}). An explicit computation of the $g=2$ case, as well as predictions at the orbifold point, will be left to section \ref{g=2}.

\subsubsection{$g>1$ and modular forms}
Let us specialize the recursion to the case of local $\mathbb{F}_2$.  The Hori-Vafa differential (\ref{differpq2}) reads, in the $(a_3,a_4)$ patch,
\beq
\label{differ22}
d\lambda_{2,2}(u)= \log\left(\frac{P_2(u) \pm Y(u)}{2} \right)\frac{du}{u}
\eeq
where
$$P_2(u)=a_3+a_4 u+ u^2, \qquad Y(u)=\sqrt{P_2^2(u)-4}$$
and the $\Gamma_{2,2}$ family can be written in the $\mathbb{Z}_2$ symmetric form (\ref{g04curve}) as a two-fold branched covering of the compactified $u$-plane
\beq Y^2 = (u-b_1)(u-b_2)(u-b_3)(u-b_4)=(\hat u^2-c_1^2)(\hat u^2-c_2^2) \label{defY}\eeq
thanks to (\ref{branchf2hom}) and (\ref{g04curve}) and having defined $\hat u=u+a_4/2$. We have first of all that
\beq
\label{thxh}
d\lambda(u)-d\lambda(\bar u)=2 M(u) Y(u) d u
\eeq
where the so-called ``moment function'' $M(u)$ is given, after using the fact that $\log({P+Y})-\log({P-Y})= 2 \tanh^{-1}{(Y/P)}$, as
\beq
M(u)={1\over u Y(u)} \tanh^{-1} \biggl[ { {Y(u)} \over P_2(u)} \biggr],
\label{defM}
\eeq
Moreover, the one form $dE(p,q)$ can be written as \cite{Eynard:2004mh}
\beq
\label{CjLambda}
d E_{w}(u)={1\over 2} {Y(w)\over Y(u)}\,\left(
{1\over u-w}-L C(w)
\right)\, d u
\eeq
where
\beq\label{defCj}
C(w):={1\over 2\pi i }\,\oint_{A} {d u \over {Y(u)}}\,{1\over u-w}, \qquad L^{-1}:= {1\over 2\pi i }\,\oint_{A} {d u \over {Y(u)}}
\eeq
We have assumed here that $w$ stays outside the contour $A$; when $w$ lies inside the contour $A$, $C(w)$ in (\ref{CjLambda}) should be replaced by its regularized version
\beq
\label{regc}
C^{\rm reg}(w)=C(w)-\frac{1}{Y(w)}
\eeq
%
%
%
%
Since $\Gamma_{2,2}$ is elliptic, it is possible to find closed form expressions for $C(u)$, $C_{reg}(u)$, $B(u,w)$ and $L$. We have 
\beq
C(u)= {2 (b_2-b_3)\over \pi (u-b_3) (u-b_2) {\sqrt{(b_1-b_3)(b_2-b_4)}}} 
\biggl[  \Pi(n_4, k) + \frac{u-b_2}{b_2-b_3} K(k)\biggr]
\eeq
\beq
C^{\rm reg}(u)=  {2 (b_3-b_2)\over \pi (u-b_3) (u-b_2) {\sqrt{(b_1-b_3)(b_2-b_4)}}} 
\biggl[  \Pi(n_1, k) + \frac{u-b_3}{b_3-b_2} K(k)\biggr] 
\eeq
\beq L^{-1} = \frac{2}{\sqrt{(b_1-b_3)(b_2-b_4)}}K\left[\frac{(b_1-b_2)(b_3-b_4)}{(b_1-b_3)(b_2-b_4)}\right]  \label{defL}\eeq
\bea B(u,w)&=&\frac{1}{Y(u)}\left[\frac{Y^2(u)}{2 Y(w) (u-w)^2}+\frac{(Y^2)'(u)}{4 Y(w) (w-u)}+\frac{A(u)}{4 Y(w)} \right]\nonumber \\ &+& \frac{1}{2(u-w)^2}
\label{bergmann}
\eea
where 
\beq
k ={(b_1-b_2)(b_3-b_4)\over(b_1-b_3)(b_2-b_4)}, \qquad n_4= {(b_2-b_1)(u-b_3) \over (b_3-b_1)(u-b_2)}, \qquad n_1= {(b_4-b_3)(u-b_2) \over (b_4-b_2)(u-b_3)},
\label{ellmodulus}
\eeq
\beq A(u)=(u-b_1)(u-b_2)+(u-b_3)(u-b_4) +(b_1-b_3)(b_2-b_4)\frac{E(k)}{K(k)} \label{defA} \eeq
and $K(k)$, $E(k)$ and $\Pi(n, k)$ are the complete elliptic integrals of the first, second and third kind respectively.
%
%
\\

\noindent With these ingredients one can compute the residues as required in (\ref{rec1}). Given that $d E_{q}(p)/(d\lambda(q)-d\lambda(\bar q))$, as a function of $q$, 
is regular at the branch-points, all residues appearing in (\ref{rec1}) will be linear combinations of the following {\it kernel differentials}
\bea
\label{kd}
\chi_i^{(n)}(p)&=&{\rm Res}_{q=x_i} \biggl(  \frac{d E_q (p)}{d\lambda(q)-d\lambda(\bar q)} {1\over (q-x_i)^n} \biggr) \nonumber \\
&=&{1\over  (n-1)!} {1\over  {Y(p)}} {d^{n-1} \over d q^{n-1}} \Biggl[{1\over 2 M(q) }\biggl( {1\over p-q} -L C(q)\biggl) \Biggr]_{q=x_i}
\eea
In (\ref{kd}), $C(p)$ should be replaced by $C_{\rm reg}(p)$ when $i=1,2$. 
%
\\

Let us then explicitly display the quasi-modular structure of the free energies $\mathcal{F}_g(a_3,\tau)$. We claim that the holomorphic limit of the $1$-parameter examples (\ref{fgeisen}) 
$$\mathcal{F}_g(\tau) = C^{2g-2}(\tau) \sum_{k=1}^n E^k_2(\tau)c^{(g)}_k(\tau) + C^{2g-2}(\tau) c^{(g)}_0(\tau) $$
gets replaced here by
\beq
\mathcal{F}_g(a_3, \tau) = C^{2g-2}(a_3, \tau) \sum_{k=1}^n E^k_2(\tau)c^{(g)}_k(a_3, \tau) + C^{2g-2}(a_3, \tau) c^{(g)}_0(a_3, \tau)
\label{fgeisen2}
\eeq
i.e., as a polynomial in the second Eisenstein series having (algebraic) functions of $a_3$ and $\theta_i(\tau)$, $i=2,3,4$ as coefficients; moreover, these coefficients are completely determined in closed form from (\ref{rec1}).  Let us show in detail how this happens in general, leaving the concrete example of the $g=2$ case to the next section. Formulae (\ref{rec2}), (\ref{rec1}), (\ref{bergmann}) and (\ref{kd}) imply that the final answer will be a polynomial in the following five objects
\beq
\label{objrec}
M_i^{(n)}, \qquad \phi^{(n)}_i, \qquad A^{(n)}_i, \qquad \left(\frac{1}{Y}\right)_i^{(n)}, \qquad \mathcal{C}^{(n)}_i
\eeq
where, for a function $f(x)$ with meromorphic square $f^2(x)$, we denote with $f^{(n)}_i$ the $(n+1)$-th coefficient in a Laurent expansion of $f(x)$ around $b_i$
\beq
f(x) = \sum_{n=-N_i}^\infty \frac{f_i^{(n+N_i)}}{(p-b_i)^{n/2}}
\eeq
and have defined
\beq
\mathcal{C}^{(n)}_i=\left\{\begin{array}{cc} C^{(n)}_{{\rm reg}, i} & \mathrm{for }\quad  i=1,2 \\ C^{(n)}_i & \mathrm{for } \quad i=3,4\end{array}\right.\eeq
Of the five building blocks in (\ref{objrec}),  $M_i^{(n)}$ and $\phi^{(n)}_i$ are the ones which are computed most elementarily from (\ref{differ22}), (\ref{philambda}) and (\ref{defM}), the result being in any case an algebraic function of $(a_3, a_4)$. 
When re-expressed in modular form, we can actually say more about them: we have that the $a_3$-dependence in $M_i^{(n)}(a_3,\tau)$ and $\phi^{(n)}_i(a_3,\tau)$ is constrained to come only through $a_4(a_3,\tau)$ as written in formula (\ref{a4mod}). Indeed, from (\ref{branchf2hom}), (\ref{branchf2modular}) we have that the branch points $b_i$ have the form
\beq
-\frac{a_4}{2} \pm c_1(\tau), \qquad -\frac{a_4}{2} \pm c_2(\tau)
\eeq
and therefore depend on $a_3$ only through $a_4(a_3,\tau)$. Moreover, since $P_2(b_i)=2$ and the derivatives of $P_2(u)$ do not depend explicitly on $a_3$, we have that the $a_3$ dependence in $\mathcal{F}_g$ as obtained from the recursion may only come through $a_4(a_3,\tau)$.
Notice moreover that these are the only pieces bringing a dependence on the additional $a_3$ variable: all the others do not depend on the form of the differential (\ref{differ22}), and are functions only of differences of branch points $b_i$. This means in particular that they only depend on the variables $c_1$ and $c_2$ introduced in (\ref{branchf2hom}) and whose modular expressions we already found in (\ref{branchf2modular})! This is immediate to see for  $A^{(n)}_i$ and $(1/Y)_i^{(n)}$ from formulae (\ref{defY}) and (\ref{defA}). The case of $\mathcal{C}^{(n)}_i$ is just slightly more complicated, but it is worth describing in detail for the discussion to come. For $n=1$, we need the first derivative of $\Pi(x,y)$ with respect to $x$:
$$ \partial_x \Pi(x,y)= \frac{x E(y)+(y-x) K(y)+\left(x^2-y\right) \Pi (x,y)}{2
   (x-1) x (y-x)}
$$
The above formula implies that 
\beq
\partial^{(n)}_x \Pi(x,y)=A_n(x,y) K(y) +  B_n(x,y) E(y)+ C_n(x,y) \Pi(x,y)
\label{derPi}
\eeq
where $A_n$, $B_n$ and $C_n$ are rational functions of $x$ and $y$. From (\ref{ellmodulus}), to compute $\mathcal{C}^{(n)}_i$, we need to evaluate these expressions when $n_1$ (resp. $n_4$) equals either 0 or $k$. But using
\beq
 \Pi(0,y)=K(y), \qquad \Pi(y,y)=\frac{E(y)}{1-y}
\eeq
we conclude that 
\beq
\mathcal{C}^{(n)}_i=R^{(n)}_1(c_1,c_2) K(k) + R^{(n)}_2(c_1,c_2) E(k)
\eeq
for two sequences of rational functions $R^{(n)}_i$. We now make the following basic observation: by (\ref{kd}), $\mathcal{C}^{(n)}_i$ always appears multiplied by $L$ in the recursion. By (\ref{defL})
\beq
L \mathcal{C}^{(n)}_i=\tilde R^{(n)}_1(c_1,c_2) + \tilde R^{(n)}_2(c_1,c_2) \frac{E(k)}{K(k)}
\label{LC}
\eeq
This last observation allows us to collect all the pieces together and state the following. By (\ref{rec1}) and (\ref{rec2}) we have that $\mathcal{F}_g(a_3, \tau)$ is a polynomial in $M_i^{(n)}$, $\phi^{(n)}_i$, $A^{(n)}_i$, $(1/Y)_i^{(n)}$, $\mathcal{C}^{(n)}_i$, and moreover the whole discussion above as well as formulae (\ref{defA}) and (\ref{LC}) imply that this takes the form of a polynomial in $W(\tau):=E(k)/K(k)$\beq
\mathcal{F}_g(a_3, \tau) = \sum_{k=0}^n W^k(\tau) h^{(g)}_k(a_3, \tau)
\eeq
with coefficients $h^{(g)}_k(a_3, \tau)$ in the ring of weight zero modular forms of $\Gamma(2)$, parametrically depending on $a_3$. \\
To conclude, we can exploit the fact that \cite{sloane}
\beq
E(k)K(k)=\left(\frac{\pi}{2}\right)^2 \frac{4 E_2(2\tau)-E_2(\tau)}{3} 
\eeq
and that from (\ref{branchf2modular})  and (\ref{Kmod})
\beq
K(k)=\frac{\pi}{2} \theta_3(\tau)\theta_4(\tau)
\eeq
where we have used the fact that in our case
$$K(k)=\sqrt{\frac{c_2}{c_1}}K\left(1-\frac{c_2^2}{c_1^2}\right) $$
as the reader can easily check. Moreover, the second Eisenstein series satisfies the duplication formula
\beq
E_2(2\tau)=\frac{E_2(\tau)}{2}+\frac{\theta_4^4(\tau)+\theta_3^4(\tau)}{4}
\eeq
Therefore,
\beq
W(\tau)=\frac{1}{3 \theta_4^2(\tau)\theta_3^2(\tau)} \left(E_2(\tau)+\theta_3^4(\tau) + \theta_4^4(\tau)\right)
\eeq
This proves (\ref{fgeisen2}). \\


\subsubsection{The $g=2$ case in detail}
\label{g=2}
Let us complete the discussion of this section by presenting the explicit formulae for the genus 2 case.  By (\ref{rec2}) and (\ref{rec1}), we need the complete expression of $W_2^{(0)}$, $W_3^{(0)}$, $W_1^{(1)}$, $W_2^{(1)}$ and $W_1^{(2)}$. The first three were computed in \cite{Bouchard:2007ys} and are given by 
\bea
W_2^{(0)}(p_1,p_2)&=& B(p_1,p_2) \nonumber \\
W_3^{(0)}(p_1, p_2, p_3)&=&{1\over 2} \sum_{i=1}^{4} M^2(b_i) (Y^2)'(b_i) \chi^{(1)}_i(p_1) \chi^{(1)}_i(p_2) \chi^{(1)}_i(p_3), \\
W_1^{(1)}(p)&=&{1\over 16} \sum_{i=1}^{4} \chi^{(2)}_i(p) +{1\over 8} \sum_{i=1}^{4} \biggl( \frac{2 A(b_i)}{(Y^2)'(b_i)} -\sum_{j\not=i} {1\over b_i -b_j} \biggr) \chi_i^{(1)}(p), \nonumber 
\label{w03w11}
\eea
$W_2^{(1)}$ is then given from (\ref{rec1}) as
\beq
W_2^{(1)}(p,p_1)= \sum_{b_i}  \underset{q=b_i}{\rm Res~} {d E_{q}(p) \over d\lambda(q) - d\lambda(\bar q)} \Big ( W^{(0)}_{3} (q, \bar q, p_1) +  2 W^{(1)}_1(q) W^{(0)}_2 (\bar q, p_1) \Big)
\eeq
A very lengthy, but straightforward computation leads us to
\beq
W_2^{(1)}(p,q) =-\frac{1}{8}\sum_{i=1}^4\left[ A_i(q)\chi_i^{(3)}(p) + B_i(q)\chi_i^{(2)}(p)+C_i(q)\chi_i^{(1)}(p) + \sum_{j\neq i} D_{ij}(q)\chi_i^{(1)}(p)\right]
\eeq
For the sake of notational brevity, we spare to the reader the very long expressions of the rational functions $A_i(q)$, $B_i(q)$, $C_i(q)$ and $D_{ij}(q)$. They involve $M_i^{(n)}$,  $A^{(n)}_i$, $(1/Y)_i^{(n)}$ and $\mathcal{C}^{(n)}_i$ up to the third order in a Taylor-Laurent expansion around the branch points. \\

\noindent The next step is given by 
\beq
W_1^{(2)}(p)= \sum_{b_i}  \underset{q=b_i}{\rm Res~} {d E_{q}(p) \over d\lambda(q) - d\lambda(\bar q)} \Big ( W^{(1)}_{2} (q, \bar q) +   W^{(1)}_1(q) W^{(1)}_1 (\bar q) \Big)
\eeq
The pole structure of $A_i(q)$, $B_i(q)$, $C_i(q)$ and $D_{ij}(q)$ dictates for $W_1^{(2)}(p)$ the following linear expression in terms of kernel differentials
\beq
W_1^{(2)}(p) = \sum_{n=1}^5\sum_{i=1}^4 E^{(n)}_i \chi_i^{(n)}(p)  
\eeq
for some (very complicated) coefficients $E_i^{(n)}$. The recursion is finalized for $g=2$ by (\ref{rec2})
\beq \mathcal{F}_2 = -{1 \over 2} \sum_{b_i} \underset{p=b_i}{\rm Res~} \phi(p) W_1^{(2)}(p) \label{F2} \eeq
It is useful to collect together terms involving the same powers of $W(\tau)$. Taking the residues in (\ref{F2}) yields\footnote{It must be noticed that, in order to match exactly the asymptotics of the Gromov-Witten expansion at large radius, we have to subtract from (\ref{F2}) a constant term in $\tau$, namely, a rational function of $a_3$ of the form $\frac{a_3^2-10}{1440 \left(a_3^2-4\right)}$. It would be interesting to investigate the origin of this discrepancy further.
}
\beq
\sum_{n=0}^3 h^{(2)}_n(a_3, \tau) W^n(\tau)
\label{f2mod}
\eeq
where
\bea
h^{(2)}_3(a_3, \tau) &=& \frac{5 a_4^2(a_3, \tau) \theta_2^4(\tau)}{24576  \theta_3^2(\tau) \theta_4^2(\tau)} \nonumber \\
h^{(2)}_2(a_3, \tau) &=& \frac{1}{1024} -\frac{a_4(a_3,\tau)^2}{49152 \theta_3(\tau)^6 \theta_4(\tau)^4} \Bigg[\theta_2(\tau)^4 (15 \theta_4(\tau)^6+16 \theta_3(\tau)^2 \theta_4(\tau)^4 \nonumber \\ &+& \theta_2(\tau)^4 \left(8    \theta_3(\tau)^2+15 \theta_4(\tau)^2\right)) \Bigg]\nonumber \\
h^{(2)}_1(a_3, \tau) &=& -\frac{\left(\theta_2(\tau)^4+2 \theta_4(\tau)^4+3 \theta_3(\tau)^2 \theta_4(\tau)^2\right)}{3072
  \theta_4(\tau)^2 \theta_3(\tau)^2} + \frac{a_4^2(a_3,\tau)}{294192}\Bigg[\frac{13\theta_2(\tau)^{12}}{\theta_3(\tau)^6\theta_4(\tau)^6} \nonumber \\ 
&+& \frac{91\theta_2(\tau)^8}{\theta_3(\tau)^6 \theta_4(\tau)^2} + 
\frac{48\theta_2(\tau)^8}{\theta_3(\tau)^4\theta_4(\tau)^4}+\frac{91\theta_4(\tau)^2
  \theta_2(\tau)^4}{\theta_3(\tau)^6}+\frac{96\theta_2(\tau)^4}{\theta_3(\tau)^4}\Bigg] \nonumber \\
h^{(2)}_0(a_3, \tau) &=& \frac{1}{61440}\left(\frac{1}{a_3+2}-\frac{1}{a_3-2}\right)+\frac{\theta_2(\tau)^8-5 \theta_3(\tau)^2 \theta_2(\tau)^4+10 \theta_3(\tau)^6}{30720 \theta_3(\tau)^4 \theta_4(\tau)^4} \nonumber \\
&+& \frac{a_4^2(a_3,\tau) \theta_2(\tau)^4}{2949120} \Bigg[12 \left(\frac{\theta_2(\tau)^4}{\theta_3(\tau)^8}-\frac{\theta_2(\tau)^4}{\theta_4(\tau)^8}\right)-\frac{65 \theta_4(\tau)^2}{\theta_3(\tau)^6}-\frac{175}{\theta_3(\tau)^2 \theta_4(\tau)^2} \nonumber \\ &-& \frac{311}{2
   \theta_3(\tau)^4}-\frac{311}{2 \theta_4(\tau)^4}-\frac{65 \theta_3(\tau)^2}{\theta_4(\tau)^6}\Bigg]+\frac{17}{46080}
\eea 
Plugging in the expression (\ref{qmod}) of the modular parameter $q$ in exponentiated flat coordinates reproduces as expected the topological vertex expansion at large radius
\bea
\mathcal{F}_2^{LR}(Q_B,Q_F) &=&
\left(-\frac{1}{120}-\frac{Q_{B}}{120}\right)
   Q_{F}+\left(-\frac{1}{60}-\frac{Q_{B}}{60}-\frac{Q_{B}^2}{60}\right)
   Q_{F}^2 \nonumber \\ &+& \left(-\frac{1}{40}-\frac{Q_{B}}{40}-\frac{Q_{B}^2}{40}-\frac{Q_{B}^3}{40}\right)
   Q_{F}^3+\left(-\frac{1}{30}-\frac{Q_{B}}{30}-\frac{Q_{B}^2}{6}-\frac{Q_{B}^3}{30}\right) Q_{F}^4 \nonumber \\
&+&\left(-\frac{1}{24}-\frac{Q_{B}}{24}-\frac{299 Q_{B}^2}{24}-\frac{299Q_{B}^3}{24}\right) Q_{F}^5+
+O\left(Q_{F}^6\right)
\eea
Finally, we can use (\ref{f2mod}) to make predictions for genus 2 orbifold Gromov-Witten invariants of $\mathbb{C}^3/\mathbb{Z}_4$  by using the Feynman expansion method of \cite{bcov, Aganagic:2006wq} as we did for the genus 1 free energy; the same result would be obtained by analytically continuing the holomorphic ambiguity $h^{(2)}_0(a_3, \tau)$ and taking the holomorphic limit of the physical amplitude directly at the orbifold point (see \cite{vinceproc} for a detailed description of this method). The results are shown in table \ref{genus2OGW}.\footnote{While the final version of this paper was under completion, a preprint appeared \cite{Alim:2008kp} where the same results have been obtained following a different method.}
\begin{table}[t]
\centering
\begin{tabular}{|c|ccccccc|}
\hline
 & $m$ & 0 & 2 & 4 & 6 & 8 & 10\\
\hline
$n$ & &  & & & & &\\
0 & & $-\frac{1}{960}$ & 0 & $-\frac{61}{30720}$ & 0 & $-\frac{9023}{81920}$ & 0 \\
1 & & 0 & $\frac{41}{46080}$ & 0 & $\frac{6061}{245760}$ & 0 & $\frac{36213661}{7864320}$ \\
2 & & $-\frac{7}{7680}$ & 0 & $-\frac{647}{92160}$ & 0 & $-\frac{1066027}{1310720}$ & 0 \\
3 & & 0 & $\frac{257}{92160}$ & 0 & $\frac{168049}{983040}$ & 0 & $\frac{887800477}{15728640}$ \\
4 & & $-\frac{11}{5120}$ & 0 & $-\frac{65819}{1474560}$ & 0 & $-\frac{18530321}{1966080}$ & 0 \\
5 & & 0 & $\frac{23227}{1474560}$ & 0 & $\frac{43685551}{23592960}$ & 0 & $\frac{62155559923}{62914560}$ \\
6 & & $-\frac{2479}{245760}$ & 0 & $-\frac{437953}{983040}$ & 0 & $-\frac{9817250341}{62914560}$ & 0 \\
7 & & 0 & $\frac{418609}{2949120}$ & 0 & $\frac{452348269}{15728640}$ & 0 & $\frac{5851085490887}{251658240}$ \\
8 & & $-\frac{19343}{245760}$ & 0 & $-\frac{303139073}{47185920}$ & 0 & $-\frac{438364727389}{125829120}$ & 0 \\
9 & & 0 & $\frac{1380551}{737280}$ & 0 & $\frac{25384681949}{41943040}$ & 0 & $\frac{355405937648809}{503316480}$ \\
10 & & $-\frac{604199}{655360}$ & 0 & $-\frac{2982122587}{23592960}$ & 0 & $-\frac{16896151842371}{167772160}$ & 0 \\
11 & & 0 & $\frac{200852963}{5898240}$ & 0 & $\frac{25012290702059}{1509949440}$ & 0 & $\frac{54049855936801961}{2013265920}$ \\
12 & & $-\frac{59566853}{3932160}$ & 0 & $-\frac{818897894611}{251658240}$ & 0 & $-\frac{1840152188554961}{503316480}$ & 0 \\
& & & & & & &
\\
\hline
\end{tabular}
\caption{Genus two orbifold Gromov-Witten invariants $N^{orb}_{2,(m,n)}$ of $[ \mathbb{C}^3/\mathbb{Z}_4 ]$}
\label{genus2OGW}
\end{table}

\section{Conclusions and outlook}

In this paper we have proposed an approach to the study of $A$-model topological amplitudes which yields exact results in $\alpha'$
and as such applies to the full moduli space, including orbifold and conifold divisors, of closed and open strings 
on a large class of toric Calabi-Yau threefolds. 
One of the main virtues of this approach is that it provides us with a closed expression for the (derivatives of the) periods 
of the mirror curve, considerably simplifying the study of their analytic continuation in the various patches and of the
modular properties of the Gromov-Witten generating functional.
The local geometries that we have analyzed arise from the 
minimal resolution of $Y^{p,q}$ singularities. 
The general procedure to compute topological string amplitudes, outlined in Section 4, is based on the correspondence with five-dimensional
gauge theories and the associated Seiberg-Witten curves; it has been fully exploited in Section 6 for the case $p=2$, and
used in particular to predict open and closed orbifold Gromov-Witten invariants of $\mathbb{C}^3/\mathbb{Z}_4$ also at higher genus. 

Of course our strategy is completely general and could be adopted with no changes, though becoming technically more involved, 
to compute amplitudes for $p>2$; moreover, it can be used to get some qualitative information about the behavior
of the B-model moduli space, which for these cases displays a richer set of phenomena.  
Indeed, the mirror curves have higher genera and can be subject to more general degeneration limits, for example when the neck
connecting two handles becomes infinitely long. In the underlying four-dimensional gauge theory
this limit has been recognized as a new superconformal phase \cite{Argyres:1995jj}; it would be interesting
to explore its interpretation in the topological string moduli space.

The computations of Section \ref{conti} have been based on extensively exploiting the holomorphic properties (\ref{holdiffpq}) of the $B$-model $1-$differential, which came out by appealing to the relation with gauge theories and integrable systems. On the gauge theory side, one is able to obtain the Seiberg-Witten curve and the related differential in a suitable
semiclassical limit involving a large number of instantons \cite{okounkov}. 
The considerations above suggest to reinterpret 
the transition to the mirror and (\ref{holdiffpq}) at the string
theoretical level in terms of a semiclassical geometry in $g_s\to 0$ which resums a large number of world-sheet instantons.

Some remarks are in order concerning the relation with integrable systems. First of all, as we have discussed in Section
3.3, the mirror geometry for resolved $Y^{p,p}$ singularities can be realized as a fibration over the spectral curve
of the relativistic $A_{p-1}$ Toda chain. Actually our results for generic $Y^{p,q}$ singularities seems to indicate
the existence of a larger class of integrable systems: it would be interesting to understand this better and see what 
kind of deformations of the Toda chain are associated to the $q$ parameter.
Moreover, the existence of a set of holomorphic differentials like (\ref{holdiffpq}) could be recognized as a signal
of a relation with integrable hierarchies. More precisely, one could expect that a suitable generalization
of the topological string prepotential - possibly including gravitational descendants - could be interpreted in terms of
a Whitham deformation of the integrable system. This would correspond to an
``uplift'' to topological strings of similar notions developed in \cite{Nakatsu:1995bz} for four-dimensional Seiberg-Witten theory. 

 
As a final comment, we might wonder how much of what we have learned might be extended to other cases. 
Moving beyond $Y^{p,q}$, it is in fact straightforward to show that holomorphicity of (derivatives of) the differential can be shown exactly as for the $Y^{p,q}$ family, at least in the case in which the mirror curve is hyperelliptic\footnote{For example, the canonical bundle over the second Del Pezzo $dP_2$ falls into this category, though not being part of the $Y^{p,q}$ class.}; at a pictorial level, this class coincides with those toric $CY$ whose toric diagram is contained into a vertical strip of width $2$, modulo $SL(2,\mathbb{Z})$ transformations. 
Our methods thus continue to hold and apply with no modification for this more general family as well;
it would be very interesting to investigate the possibility to generalize our approach to all toric Calabi-Yau
three-folds.



\hyphenation{Ma-no-la-che}
\vskip 1truecm
 \noindent {\large{\bf Acknowledgments}}
\\
We would like to thank G.~Bonelli, H.~L.~Chang, B.~Dubrovin, B.~Fantechi, C.~Manolache, F.~Nironi, S.~Pasquetti, E.~Scheidegger for 
useful conversations, and we are particularly grateful to Tom Coates 
for enlightening discussions of Section 5.2. 
A.B. acknowledges I.~Krichever, O.~Ragnisco, S.~Ruijsenaars, N.~Temme and especially Ernst D.~Krupnikov for kind email correspondence, and
V.~Bouchard, M.~Mari\~no and S.~Pasquetti for fruitful and stimulating discussions during the final phase of this project. 
The present work is partially
supported by the European Science Foundation Programme ``Methods
of Integrable Systems, Geometry, Applied Mathematics'' (MISGAM) and
Marie Curie RTN ``European Network in Geometry, Mathematical Physics
and Applications'' (ENIGMA). 

\vskip 1truecm
 \noindent {\Large{\bf Appendix}}
\begin{appendix}
\section{Euler integral representations, analytic continuation and generalized hypergeometric functions}
\label{analytic}
As we pointed out in section \ref{sezlauric}, another important feature of our formalism is the fact that we can work directly with an Euler-type integral representation for the periods. We will focus here in the case $p=q=2$, but the strategy is completely general and 
computationally feasible as long as $x_i$ is algebraically related to $a_i$.\\
For $p=q=2$, the derivatives of the periods have the simple form (\ref{PiAF2}), (\ref{PiBF2}). Using the standard Euler integral representation for the complete elliptic integral $K(x)$ 
\beq
2 K(x)=\int_0^1 \frac{d\theta}{\sqrt{\theta}\sqrt{1-\theta}} \frac{1}{\sqrt{1- x\theta}}
\eeq
we can integrate back $a_4$ and get

\bea
\label{PiAEuler}
\Pi_A(a_i) &=& \int_{0}^1 \frac{2 a_{5} d\theta}{\sqrt{\theta} \sqrt{1-\theta}} \log \left[a_{4}+\sqrt{c_1^2 + (c_2^2-c_1^2)    (1-\theta)}\right] \\
\Pi_B(a_i) &=& 4\int_{0}^1 \frac{2a_{5} d\theta}{\sqrt{1-\theta^2} \theta} 
\log\left[\frac{1}{\sqrt{c_1^2-c_2^2}}\left(a_4 \theta  + \sqrt{(c_1^2-c_2^2) + c_2 \theta^2}\right)\right] 
\label{PiBEuler}
\eea
where the constant factors in $a_4$ are introduced as a constant of integration in order to satisfy (\ref{GKZeq}). Formulae (\ref{PiAEuler}), (\ref{PiBEuler}) then yield simple and globally valid expressions for the periods and significantly ease the task of finding their analytic continuation from patch to patch. For small $a_4$, we can simply expand the integrand and integrate term by term. For large $a_4$ $\Pi_A$ has the following asymptotic behavior

\beq
\label{singlelog}
\Pi_A = 2 a_{5} \log \left(2 a_4\right) -2 \left(a_{3} a_{5}^2\right) \left(\frac{1}{a_{4}}\right)^2+\left(-3
   a_{3}^2 a_{5}^3-6 a_{1} a_{2} a_{5}^3\right) \left(\frac{1}{a_{4}}\right)^4+O\left(\frac{1}{a_{4}}\right)^5
\eeq
but an expansion for $\Pi_B$ is much harder to find. The leading order term can still be extracted, for example in the $a_2=a_3=a_5=1$ patch using
\beq
\int_0^1 \log{\left[\theta a + \sqrt{1+\left(b+\frac{a^2}{4}\right) \theta^2} \right]} \frac{d\theta}{\sqrt{1-\theta^2}\theta} = 2 Li_2(-1-a) + O(\log{a})
\eeq
which gives

\beq
\label{doublelog}
\Pi_B = 4 \left(\log \left(\frac{1}{2
   \sqrt[4]{a_{1}}}\right)-\log
   \left(\frac{1}{a_{4}}\right)\right)^2 + O\left(\log{a_4}\right)
\eeq
Single and double logarithmic behaviors as in (\ref{singlelog}, \ref{doublelog}) are  characteristic of the large radius patch in the moduli space, which as we will see will be given precisely by $a_4\to \infty$ (and $a_1 \to 0$).\\

Lastly, a nice fact to notice is that the periods for this particular case take the form of known generalized hypergeometric functions of two variables. For example we have that, modulo $a_4$ independent terms, the $A$ period can be written as

\beq
\Pi_A = \frac{\pi}{4} \log{c_1}+\frac{\pi}{4}\left(\frac{a_4^2}{c_1}-c_1\right)F^{1, 2, 2}_{1, 1, 1}\left[\begin{array}{c|c|c}1 & \frac{3}{2}, \quad  1 & \frac{1}{2} , \quad\frac{1}{2}\\ 2 &  2 &  1\end{array}\Bigg| c_1\left(1-\frac{a_4^2}{c_1^2}\right), c_1\left(1-\frac{c_2^2}{c_1^2}\right)\right]
\eeq
in terms of the Kamp\'e de F\'eriet\footnote{See Eric Weinstein, {\it ``Kamp\'e de F\'eriet Function''}, \texttt{http://mathworld.wolfram.com/KampedeFerietFunction.html}, or \cite{extonfun, extonint} for a more detailed account on such functions.} hypergeometric function of two variables.

\section{Lauricella functions}
\label{applauric}
We collect here a number of properties and useful formulae for Lauricella's $F_D^{(n)}$ functions. The interested reader might want to look at \cite{extonfun} for a detailed discussion of this topic.

\subsection{Definition}
The usual power series definition of Lauricella $F_D^{(n)}$ of $n$ complex variables is
\bea\label{power}
& & F_D^{(n)}(a,b_1,\ldots,b_n;c;x_1,\ldots,x_n) \nonumber \\
	&=&\sum_{m_1=0}^\infty\cdots \sum_{m_n=0}^\infty
	\frac{(a)_{m_1+\cdots+m_n}(b_1)_{m_1}\cdots(b_n)_{m_n}}{(c)_{m1+\cdots+m_n}m_1!\cdots m_n!}
	\; x_1^{m_1}\cdots x_n^{m_n},
\eea
whenever $| x_1|,\ldots,| x_n | < 1$. For $n=1$ this is nothing but Gauss' hypergeometric function $_2F_1(a,b;c;x)$; for $n=2$ is boils down to Appell's $F_1(a,b,c;d;x,y)$. It also satisfies the following system of $PDE$'s, which generalizes the $n=1$ hypergeometric equation

\bea
\label{systfd}
a b_j F_D &= & x_j (1-x_j) \frac{\partial^2 F_D}{\partial x_j^2}+(1-x_j) \sum_{k\neq j} x_k \frac{\partial^2 F_D}{\partial x_k \partial x_j} + [c-(a+b_j+a) x_j] \frac{\partial F_D}{\partial x_j} \nonumber \\ &- &  b_j \sum_{\neq j} x_k \frac{\partial F_D}{\partial x_k} \qquad \quad j=1, \dots, n
\eea
The system (\ref{systfd}) has regular singular points when

\beq
x_i=0,1,\infty \quad \hbox{and} \quad x_i=x_j \qquad i=1,\dots,n, j\neq i
\eeq
The number of intersecting singular submanifolds in correspondence of the generic singular point

\beq
(x_1, \dots, x_n)= (\underbrace{0,\dots, 0}_{p}, \underbrace{1\dots,1}_{q},\underbrace{\infty,\dots,\infty}_{n-p-q})\\
\eeq
is

$$\left(\begin{array}{c}p+1 \\ 2\end{array}\right) \left(\begin{array}{c}q+1 \\ 2\end{array}\right) \left(\begin{array}{c}n-p-q+1 \\ 2\end{array}\right) $$

In contrast with the well-known $n=1$ case, typically the Lauricella system does not close under analytic continuation around a singular point. As explained in \cite{extonfun}, a complete set of solutions of the $F_D^{n}$ system (\ref{systfd}) away from the region of convergence $|x_i|<1$ involves a larger set of functions, namely Exton's $C^{k}_n$ and $D_{(n)}^{p,q}$. We will report here a number of analytic continuation formulae valid for generic $n$, and refer to \cite{extonfun} for further results in this direction. See also \cite{lopez} for further developments in finding asymptotic expressions for large values of the parameters.

\subsection{Analytic continuation formulae for Lauricella $F_D$}
In the following, results on analytic continuation for $F_D$ will be expressed in terms of Exton's $C$ and $D$ functions

\beq
\begin{array}{l}
C_n^{(k)}(\{b_i\},a,a';\{x_i\})\\=\sum_{m_1,\dots m_n} \prod_i (b_i)_{m_i} (a)_{\sum_{i=k+1}^n m_{i}-\sum_{i=1}^k m_{i}}(a')_{-\sum_{i=k+1}^n m_{i}+\sum_{i=1}^k m_{i}}\prod_i \frac{x_i^{m_i}}{m_i!}
\end{array}
\eeq

\beq
\begin{array}{l}
\label{Dpower}
D_{(n)}^{p,q}(a,b_1,\ldots,b_n;c,c';x_1,\ldots,x_n)\\
	=\sum_{m_1=0}^\infty\cdots \sum_{m_n=0}^\infty
	\frac{(a)_{m_{p+1}+\cdots+m_n-m1-\cdots-m_p}(b_1)_{m_1}\cdots(b_n)_{m_n}}{(c)_{m_{q+1}+\cdots+m_n-m_1-\cdots-m_p}c'_{m_{p+1}+\cdots+m_q}m_1!\cdots m_n!}
	\; x_1^{m_1}\cdots x_n^{m_n},
\end{array}
\eeq
\begin{itemize}
\item {\bf Continuation around $(0,0,\dots,0,\infty)$}
\end{itemize}
\beq
\begin{array}{l}
F_D^{(n)}(a,b_1,\dots, b_n;c; x_1,\dots , x_n) = \\
 \Gamma\left[
\begin{array}{cc}
 c, & b_n-a \\ 
b_n, & c-a
\end{array}
\right](-x_n)^{-a}F_D^{(n)}(a,b_1, \dots, b_{n-1}, 1-c+a;1-b_n+a; \frac{x_1}{x_n}, \dots, \frac{x_{n-1}}{x_n}, \frac{1}{x_n}) \\
+ \Gamma\left[
\begin{array}{cc}
 c, & -b_n+a \\ 
a, & c-b_n
\end{array}
\right](-x_n)^{-b}C_n^{(n-1)}(b_1, \dots, b_n, 1-c+b_n;a-b_n; -x_1, \dots, -x_{n-1}, \frac{1}{x_n})
\end{array}
\eeq

\begin{itemize}
\item {\bf Continuation around $(0,0,\dots,0,1)$}
\end{itemize}
\beq
\begin{array}{l}
F_D^{(n)}(a,b_1,\dots, b_n;c; x_1,\dots , x_n) =  \Gamma\left[
\begin{array}{cc}
 c, & c-b_n-a \\ 
c-a, & c-b_n
\end{array}
\right](1-x_1)^{-b_1}\dots(1-x_{n-1})^{-b_{n-1}}  \\
\times x_n^{-b_n} C_n^{(n-1)}(b_1,\dots, b_n, 1+b_n-c;c-a-b_n; \frac{x_1}{1-x_1}, \dots, \frac{x_{n-1}}{1-x_{n-1}}, \frac{1-x_n}{x_n}) \\
+ \Gamma\left[
\begin{array}{cc}
 c, & b_n+a-c \\ 
a, & b_n
\end{array}
\right](1-x_1)^{-b_1}\dots(1-x_{n-1})^{-b_{n-1}}(1-x_{n})^{c-a-b_{n}} \\
\times F_D^{(n)}(c-a,b_1,\dots, b_{n-1};c-a-b_n+1; \frac{1-x_n}{1-x_1},\dots , \frac{1-x_n}{1-x_{n-1}},1-x_n)
\end{array}
\eeq

\begin{itemize}
\item {\bf Continuation around $(0,0,\dots,\infty,1)$}
\end{itemize}
\beq
\begin{array}{l}
F_D^{(n)}(a,b_1,\dots, b_n;c; x_1,\dots , x_n) =  
\Gamma\left[
\begin{array}{cc}
 c, & b_n+a-c \\ 
a, & b_n
\end{array}
\right](1-x_{n})^{c-a-b_{n}}\prod_{i=1}^{n-1} (1-x_i)^{-b_i} \\
\times F_D^{(n)}(c-a,b_1,\dots, b_{n-1};c-b_1-\dots -b_n; c-a-b_n+1; \frac{1-x_n}{1-x_1},\dots , \frac{1-x_n}{1-x_{n-1}},1-x_n) \\
+\Gamma\left[
\begin{array}{ccc}
 c, & c-a-b_n, a-b_{n-1} \\ 
c-a, & c-b_{n-1}-b_n, & a
\end{array}
\right](1-x_1)^{-b_1}\dots(1-x_{n-1})^{-b_{n-1}}x_{n}^{-b_{n}}  \\
\times D_{(n)}^{1,2}(c-a-b_n,b_n, \dots, b_1; c-b_{n-1}-b_n;b_{n-1}-a+1; \frac{x_{n}-1}{x_n}, \frac{1}{1-x_{n-1}},\frac{x_{n-2}}{1-x_{n-2}} \dots, \frac{x_1}{1-x_{1}}) \\
+\Gamma\left[
\begin{array}{cc}
 c, & b_{n-1}-a \\ 
c-a, & b_{n-1}
\end{array}
\right](1-x_{n-1})^{-a} \\ \times F_D^{(n)}(a,b_1,\dots, b_{n-2};c-\sum_{i=1}^n b_i; b_n,a-b_{n-1}+1; \frac{1-x_1}{1-x_{n-1}},\dots , \frac{1-x_{n-2}}{1-x_{n-1}},\frac{1}{1-x_{n-1}},\frac{1-x_n}{1-x_{n-1}})
\end{array} 
\eeq

Notice that the formulae above are valid only for generic values of the
parameters $b_i$, $a$ and $c$. Should one be confronted with singular cases, it would be necessary to take a suitable regularization (such as $b_i\to b_i+\epsilon$) and after analytic continuation take the $\epsilon \to 0$ limit. See Appendix B in \cite{Akerblom:2004cg} for more details; suffice it here to report as an example the case $b_n=a$:

\beq
\begin{array}{l}\label{cont}
F_D^{(n)}(a,b_1,\ldots,b_{n-1},a;c;x_1,\ldots,x_n)\\
=\Gamma\left[\begin{array}{c}c\\a,c-a\end{array}\right](-x_n)^{-a}
	\sum_M\sum_{m_n=0}^\infty
	\Gamma\left[\begin{array}{c}c-a- |M| |\\c-a+| M| \end{array}\right]\frac{(a)_{| M |+m_n}(1-c+a)_{2| M|+m_n}}
	{(| M | +m_n)!m_n!}\prod_{i=1}^{n-1}\frac{(b_i)_{m_i}}{m_i!}\times\\
	\times\left(\log(-x_n)+h_{m_n}\right)\left(\frac{x_1}{x_n}\right)^{m_1}\cdots
	\left(\frac{x_{n-1}}{x_n}\right)^{m_n-1}\left(\frac{1}{x_n}\right)^{m_n}\\
+\Gamma\left[\begin{array}{c}c,c-a\\a\end{array}\right](-x_n)^{-a}\sum_M\sum_{m_n=0}^{| M |-1}
	\frac{(a)_{m_n}\Gamma(| M | - m_n)}{m_n!(c-a)_{| M |-m_n}}
	\prod_{i=1}^{n-1}\frac{(b_i)_{m_i}}{m_i!}x_1^{m_1}\cdots x_{n-1}^{m_{n-1}}\left(\frac{1}{x_n}\right)^{m_n},
\end{array}
\eeq
with
\begin{equation}
h_{m_n}=\psi(1+| M |+m_n)+\psi(1+m_n)-\psi(a+| M |+m_n)-\psi(c-a-m_n),
\end{equation}
and $M=(m_1, \dots, m_n)$ is a multindex (so that $| M |\equiv \sum_{i=1}^{m_n}m_i$).

\end{appendix}

\end{document}